\documentclass[fleqn,usenatbib]{mnras}

\usepackage{newtxtext,newtxmath}

\usepackage[T1]{fontenc}
\usepackage{ae,aecompl}

\usepackage{graphicx}	
\usepackage{amsmath}	
\usepackage{amssymb}	
\usepackage{pifont}

\newcommand{\kms}{\mbox{km s$^{-1}$}}

\newcommand{\fif}{\mbox{$\times 10^{15}$}}
\newcommand{\six}{\mbox{$\times 10^{16}$}}
\newcommand{\sev}{\mbox{$\times 10^{17}$}}
\newcommand{\eig}{\mbox{$\times 10^{18}$}}
\newcommand{\nin}{\mbox{$\times 10^{19}$}}
\newcommand{\tick}{\ding{51}}%
\newcommand{\cross}{\ding{55}}

\title[Molecular gas in the cores of BCGs]{Constraining cold accretion onto supermassive black holes: molecular gas in the cores of eight brightest cluster galaxies revealed by joint CO and CN absorption}

\author[Tom Rose et al.]{Tom Rose,$^{1}$\thanks{E-mail: thomas.d.rose@durham.ac.uk}
A. C. Edge$^{1}$, 
F. Combes$^{2}$,
M. Gaspari$^{3}$\thanks{\textit{Lyman Spitzer Jr.} Fellow.},
S. Hamer$^{4}$,
N. Nesvadba$^{5}$,\newauthor
A. B. Peck$^{6}$,
C. Sarazin$^{7}$,
G. R. Tremblay$^{8}$,
S. A. Baum$^{9,10}$,
M. N. Bremer$^{11}$,\newauthor
B. R. McNamara$^{12}$,
C. O'Dea$^{9,13}$,
J. B. R. Oonk$^{14,15,16}$,
H. Russell$^{17}$,
P. Salom\'e$^{2}$,\newauthor
M. Donahue$^{18}$,
A. C. Fabian$^{17}$, 
G. Ferland$^{19}$,
R. Mittal$^{20}$, 
A. Vantyghem$^{9}$
\\ \\
Institutions are listed at the end of the paper.
}
\date{Accepted XXX. Received YYY; in original form ZZZ}

\pubyear{2019}

\begin{document}
\label{firstpage}
\pagerange{\pageref{firstpage}--\pageref{lastpage}}
\maketitle

\begin{abstract}
To advance our understanding of the fuelling and feedback processes which power the Universe's most massive black holes, we require a significant increase in our knowledge of the molecular gas which exists in their immediate surroundings. However, the behaviour of this gas is poorly understood due to the difficulties associated with observing it directly. We report on a survey of 18 brightest cluster galaxies lying in cool cores, from which we detect molecular gas in the core regions of eight via carbon monoxide (CO), cyanide (CN) and silicon monoxide (SiO) absorption lines. These absorption lines are produced by cold molecular gas clouds which lie along the line of sight to the bright continuum sources at the galaxy centres. As such, they can be used to determine many properties of the molecular gas which may go on to fuel supermassive black hole accretion and AGN feedback mechanisms. The absorption regions detected have velocities ranging from -45 to 283 \kms\space relative to the systemic velocity of the galaxy, and have a bias for motion towards the host supermassive black hole. We find that the CN N = 0 - 1 absorption lines are typically 10 times stronger than those of CO J = 0 - 1. This is due to the higher electric dipole moment of the CN molecule, which enhances its absorption strength. In terms of molecular number density CO remains the more prevalent molecule with a ratio of \mbox{CO/CN $\sim 10$}, similar to that of nearby galaxies. Comparison of CO, CN and H\thinspace I observations for these systems shows many different combinations of these absorption lines being detected.
\end{abstract}




\section{Introduction}

Our understanding of how the molecular gas of cool-core brightest cluster galaxies behaves is largely derived from a mixture of theory \citep[e.g.][]{ODea1994,Nulsen2005,Pizzolato2005,McNamara2012,McNamara2016}, simulations \citep[e.g.][]{Gaspari2010} and emission line studies \citep[e.g.][]{Crawford1999, Edge2002,Jaffe2005,Donahue2011, Olivares2019}. Although many theoretical works hypothesise about the behaviour of this molecular gas across a wide range of spatial scales, observational studies typically focus on emission, which probes gas within relatively large collections of molecular clouds and struggles to reveal how it behaves in more compact regions. This includes areas of particular interest, such as the surroundings of the most massive supermassive black holes. As a result of this observational shortfall, there exists a significant gap in our knowledge concerning the behaviour and properties of the molecular gas surrounding active galactic nuclei (AGN). While observations have been absent at this level, simulations such as chaotic cold accretion have predicted that the large reservoirs of molecular gas we see observationally \citep[e.g. ][]{Edge2001}, exist at least in part, as a population of relatively small clouds in the few hundred parsecs around the cores of massive brightest cluster galaxies \citep[e.g.][]{Pizzolato2005, Gaspari2015, Gaspari2018}. The ensemble of molecular gas clouds which make up this reservoir are expected to undergo inelastic collisions, causing them to lose angular momentum and be funneled into the few hundred parsecs surrounding the supermassive black hole, eventually providing it with fuel. One important but currently missing observational constraint on this proposed class of AGN feedback scenarios is to determine the properties of the cooled gas such as the mass, temperature, dynamics and its origins, as well as what fraction of it ultimately becomes fuel for future outbursts from the central supermassive black hole.

A small number of recent studies of the molecular gas in the central regions of massive galaxies have begun to focus on molecular absorption, rather than emission. Such absorption line studies have two principal benefits. First, observing absorption against a bright and unresolved backlight makes it possible to study the behaviour and properties of molecular gas on much smaller scales than is achievable from emission. Second, in absorption line studies using a galaxy's bright radio core as a backlight, redshifted absorption unambiguously indicates inflow while blueshifted lines indicate outflow. In the case of emission lines, there is ambiguity as to whether the gas being traced lies in front of or behind the core of the galaxy. Despite these advantages, molecular absorption studies remain rare, with only a handful of absorbing systems having been found in this way. In terms of brightest cluster galaxies, five associated absorbers have been detected, where the absorption is observed in the spectrum of the bright radio source spatially coincident with the galaxy's supermassive black hole \citep{David2014, Tremblay2016, Ruffa2019, Rose2019, Nagai19}. A small selection of intervening absorbers have also been detected in gravitational lens systems, where absorption is observed in a galaxy separate from, but along the same line of sight as a distant and bright radio continuum source such as a quasi-stellar object \citep{Combes2008, Wiklind2018, Combes2019}.

Two of the associated absorbers detected so far have provided some indication of cold, molecular gas clouds falling towards their host galaxy's core and thus potentially going on to fuel the supermassive black hole. These observations of NGC 5044 by \citet{David2014} and of Abell 2597 by \citet{Tremblay2016} both show regions of cold molecular gas moving towards the galaxy centre at \mbox{$\sim 200-300$ \kms}. 
Additionally, \citet{Ruffa2019} and \citet{Rose2019} both show molecular gas which appears to be in stable, slightly elliptical orbits where they most likely drift in a turbulent velocity field, rather than undergoing any significant inflow or outflow. 

The molecular gas detected in these systems provides some evidence in line with theories and simulations which predict a gradual drifting of molecular clouds towards a galaxy's central supermassive black hole. However, with such a small number of detections having been made so far, it is difficult to draw concrete conclusions about the typical behaviour and properties of the molecular gas in the central regions of massive galaxies and how it interacts with the central supermassive black hole. Here we present the results of an Atacama Large Millimeter/submillimeter Array (ALMA) survey of 18 brightest cluster galaxies, all of which are extremely bright and core dominated in the radio. We find evidence of cold gas in the core regions of eight of this sample through the detection of molecular absorption lines. As well as detecting CO absorption, the sample also reveals several absorption lines of CN, a tracer of dense gas in the presence of ultraviolet radiation. There is also a detection of one SiO absorption line. Across the eight systems in which we find molecular absorption, there are 15 new individual CO, CN or SiO absorption lines detected. 

This paper is laid out as follows. In \S\ref{sec:SampleandObservations} we give details of the observations and introduce the sample, while \S\ref{sec:DataReduction} we discuss the data reduction we have carried out. In \S\ref{sec:Detections} we present the eight systems with detections of CO, CN and SiO absorption lines. In \S\ref{sec:emitting_sources} we show the sources which have CO and CN emission, but lack absorption features and in \S\ref{sec:NonDetections} we briefly discuss the sources which have no absorption or emission features. In \S\ref{sec:ColumnDensityDerivations} we derive the CO and CN column densities from the observed absorption features. In \S\ref{sec:Discussion} we discuss the significance and implications of our results and in \S\ref{sec:conclusions} we present our main conclusions. Throughout, we assume a flat $\Lambda$CDM Universe with \mbox{$H_{0}=70$ km s$^{-1}$ Mpc$^{-1}$}, \mbox{$\Omega_{M}$=0.3} and \mbox{$\Omega_{\Lambda}$=0.7}.

\begin{table*}
    \centering
    \begin{tabular}{lcccccc}
    \hline
       Source & CO(1-0) & CN-A & CN-B & SiO(3-2) & Archival CO(2-1) & Archival H\thinspace\small I\normalsize\space \\
        \hline
        Hydra-A & \tick\tick & \tick\cross & \tick\cross & - & \tick\tick & \tick\cross \\ \rule{0pt}{3.0ex} 
        S555 & \tick\tick & \tick\cross & \tick\cross & - & - &\cross\cross\\ \rule{0pt}{3.0ex} 
        Abell 2390 & \tick\cross & \tick\cross & \tick\cross & \tick\cross & - & \tick\cross \\ \rule{0pt}{3.0ex} 
        RXCJ0439.0+0520 & \tick\cross &\cross\cross&\cross\cross&\cross\cross& - &\cross\cross\\ \rule{0pt}{3.0ex} 
        Abell 1644 & \cross\tick & \tick\cross & \tick\cross & - & - & \tick\cross \\ \rule{0pt}{3.0ex} 
        NGC 5044 &\cross\cross& \tick\cross & \tick\cross & - & \tick\cross &\cross\cross\\ \rule{0pt}{3.0ex} 
        NGC 6868 & \tick\tick & \tick\cross & \tick\cross & - & - & \tick\cross \\ \rule{0pt}{3.0ex} 
        Abell 2597 & \cross\tick & \tick\cross & \tick\cross & - & \tick\tick & \tick\cross \\
        \hline \hline
        RXCJ1350.3+0940 & \cross\tick & \cross\tick & \cross\tick & - & - & \tick\cross \\ \rule{0pt}{3.0ex} 
        MACSJ1931.8-2634 & \cross\tick & - & - & - & - & - \\ \rule{0pt}{3.0ex} 
        RXCJ1603.6+1553 & \cross\tick &\cross\cross&\cross\cross& - & - & \tick\cross \\ 
        \hline\hline
        RXCJ0132.6-0804 & \cross\cross* &\cross\cross&\cross\cross&\cross\cross& - & - \\ \rule{0pt}{3.0ex} 
        MACSJ0242-2132 &\cross\cross& - &\cross\cross&\cross\cross& - & - \\ \rule{0pt}{3.0ex}  
        Abell 3112 &\cross\cross&\cross\cross&\cross\cross& - & - &\cross\cross\\ \rule{0pt}{3.0ex} 
        Abell 496 &\cross\cross&\cross\cross&\cross\cross& - & - &\cross\cross\\ \rule{0pt}{3.0ex} 
        Abell 2415 &\cross\cross&\cross\cross&\cross\cross& - & - & \tick\cross \\ \rule{0pt}{3.0ex} 
        Abell 3581 &\cross\cross&\cross\cross&\cross\cross& - & - & - \\ \rule{0pt}{3.0ex} 
        RXCJ1356-3421 &\cross\cross&\cross\cross&\cross\cross&\cross\cross& - & \tick\cross \\
        \hline
        \multicolumn{7}{c}{- Not observed} \\ 
        \multicolumn{7}{c}{\cross\cross\space Not detected in emission or absorption} \\
        \multicolumn{7}{c}{\tick\cross\space Absorption detected, emission undetected} \\ 
        \multicolumn{7}{c}{\cross\tick\space Absorption undetected, emission detected} \\ 
        \multicolumn{7}{c}{\tick\tick\space Absorption and emission detected} \\
    \end{tabular}
    \caption{For the 18 sources observed in our survey, the above table highlights the lines for which observations have been carried out and detections of emission and absorption lines have been made. We also indicate where archival CO(2-1) and H\thinspace\small I\normalsize\space observations and detections are known. The top section of the table gives these details for the sources shown in Fig. \ref{fig:CO_10_Absorbers_first_plot} and \ref{fig:CO_10_Absorbers_second_plot}, where we find $\geq 3 \sigma$ CO(0-1), CN-A and/or CN-B absorption lines. The CN-A and CN-B lines are produced when CN N = 0 - 1 absorption, which has two groupings of hyperfine structure, is observed at low spectral resolution (a more detailed description of this is given in \S\ref{sec:DataReduction}). In the middle section of the table are the sources which have clear CO(1-0)/CN-A/CN-B emission but no $\geq 3 \sigma$ absorption lines (Fig. \ref{fig:CO_10_Absorbers_third_plot}). In the lower section are the sources which do not show any $\geq 3 \sigma$ CO(1-0) or CN-A/CN-B emission and absorption along the line of sight the galaxy's continuum source. \newline *Detected in emission on scales significantly larger than the beam size.}
    \label{tab:observations_detections_table}
\end{table*}

\begin{table*}
	\centering
	\begin{tabular}{lcccccr} 

		\hline
		 & Hydra-A & S555 & Abell 2390 & RXCJ0439.0+0520 & Abell 1644 & NGC 5044\\
		\hline
		Observation date & 2018 Jul 18 & 2018 Jan 23 & 2018 Jan 07 & 2018  Jan 11 & 2018 Aug 21 & 2018 Sep 20 \\
		Integration time (s) & 2700 & 2800 & 8000 & 1300 & 2800 & 2400 \\
		CO(1-0) vel. resolution (\kms) & 2.7 & 2.6 & 3.1 & 3.0 & 2.6 & 2.5 \\
		CN vel. resolution (\kms) & 46 & 45 & 63 & 60 & 45 & 42 \\
		SiO(3-2) vel. resolution (\kms) & - & - & 54 & - & - & - \\
		Angular resolution (arcsec) & 1.63 & 0.81 & 0.37 & 0.43 & 1.97 & 0.56 \\
		Spatial Resolution (kpc) & 1.72 & 0.70 & 1.36 & 1.46 & 1.83 & 0.11 \\
		PWV (mm) & 2.85 & 2.23 & 2.12 & 2.58 & 1.39 & 0.49 \\
		FoV (arcsec) & 61.6 & 71.0 & 63.8 & 62.7 & 61.1 & 58.8 \\
		ALMA configuration & C43-1 & C43-5 & C43-6 & C43-5 & C43-3 & C43-5\\
		Maximum spacing (m) & 161 & 1400 & 2500 & 1400 & 500 & 1400 \\
		CO(1-0) noise/channel (mJy/beam) & 1.00 & 0.45 & 0.25 & 0.76 & 0.65 & 0.59  \\
		CN noise/channel (mJy/beam) & 0.16 & 0.064 & 0.030 & 0.11 & 0.10 & 0.073 \\
		SiO noise/channel (mJy/beam) & - & - & 0.064 & - & - & - \\
		115 GHz cont. flux density (mJy) & 81.5 & 12.8 & 7.7 & 72.0 &41.8 & 14.6 \\
		\hline 
		CO(2-1) channel width (\kms) &- &- &- &- &- & 1.3\\
		CO(2-1) noise per channel (mJy) &- & -&- & -&- & 0.95 \\
		\hline
	\end{tabular}

	\begin{tabular}{lccccr} 

		\hline
		 & NGC 6868 & Abell 2597 & RXCJ1350.3+0940 & MACSJ1931.8-2634 & RXCJ1603.6+1553 \\
		\hline
		Observation date & 2018 Jan 25 & 2018 Jan 02 & 2018 Sep 16 & 2018 Jan 02 & 2018 Sep 16 \\
		Integration time (s) & 5100 & 7300 & 5600 & 5300 &1500\\
		CO(1-0) vel. resolution (\kms) & 2.5 & 2.7 & 2.9 & 3.4 & 2.8 \\
		CN vel. resolution (\kms) & 42 & 48 & 53 & - & 51 \\
		Angular resolution (arcsec) & 0.81 & 0.35 & 0.66 & 0.47 & 0.68 \\
		Spatial Resolution (kpc) & 0.15 & 0.54 & 1.55 & 2.33 & 1.36 \\
		PWV (mm) & 6.52 & 1.87 & 0.66 & 3.19 & 0.82\\
		FoV (arcsec) & 58.8 & 63.3  & 66.5 & 68.3 & 65.0 \\
		ALMA configuration & C43-5 & C43-6 & C43-4 & C43-6 & C43-4 \\
		Maximum spacing (m) & 1400 & 2500 & 784 & 2500 & 784 \\
		CO(1-0) noise/channel (mJy/beam) & 0.53 & 0.34 & 0.31 & 0.24 & 0.62 \\
		CN noise/channel (mJy/beam) & 0.064 & 0.054 & 0.047 & - & 0.12 \\
		115 GHz cont. flux density (mJy) & 14.3 & 7.8 & 10.6 & 3.1 & 54.3 \\
		\hline 
		CO(2-1) channel width (\kms) & - & 4.3 & - & - & -\\
		CO(2-1) noise per channel (mJy) & -& 0.23 & - & -& - \\
		\hline
	\end{tabular}

    \caption{A summary of the ALMA observations presented in this paper, all of which were taken using ALMA band 3 and have a frequency resolution of 977 kHz. The field of view (FoV) is defined as the FWHM of the primary beam. The last two rows of the table also show the channel width and noise per channel of the archival CO(2-1) observations discussed later in \S\ref{sec:Detections} and shown in Fig. \ref{fig:CO_10_Absorbers_first_plot} and \ref{fig:CO_10_Absorbers_second_plot}.}
    \label{tab:observations_table}
\end{table*}

\section{Target sample and observed lines}
\label{sec:SampleandObservations}

The observations presented in this paper are from an ALMA Cycle 5 survey of core dominated brightest cluster galaxies with extremely high flux densities (project 2017.1.00629.S). In total, time was awarded for observations of 23 targets but three observations were not attempted and two were not sufficiently well calibrated to extract a reliable spectrum.
All 23 targets have unresolved emission at \mbox{85 - 110 GHz} of >10 mJy, so they are both bright and compact enough to probe the behaviour of cold molecular gas along very narrow, uncontaminated lines of sight. In all but one case our ALMA observations of each galaxy's radio core are unresolved. The exception to this is Abell 3112, though we see no molecular absorption in this system. For the interested reader, all of the observations presented in this paper, including all continuum images, are publicly available via the ALMA Science Archive as of September 20 2019. 

The sample of 23 brightest cluster galaxies was drawn from over 750 X-ray selected clusters with complete, multi-frequency radio coverage \citep{Hogan2015} that extends to the sub-mm for the brightest sources \citep{Hogan2015b}. Each source has either a detection at 2~mm above 7~mJy \citep{Hogan2015b}, an AT20G 20~GHz \citep{Murphy2010} detection and/or is included in the OVRO~40m 15 GHz monitoring sample \citep{Richards2011}. There are at most two sources below our declination limit ($<33^\circ$) not included in the original Cycle 5 request (Abell 2055 and Abell 2627) which meet these high frequency selection criteria but are both potentially BL~Lac dominated and so are excluded to avoid issues of source orientation \citep{Green2017}. Therefore, we are confident that the sample studied is representative and an essentially complete selection of the brightest mm-bright cores in cluster centres. All of the objects observed, 11 of which appear in the optical emission line sample of \citet{Hamer2016}, would most likely be classified as low-ionization nuclear emission-line regions (LINERs) in terms of the line widths of their optical spectra. We are not aware of any observations which would suggest any of the sample could be classified as Seyferts.

Observations were taken between 2018 January 02 and 2018 September 20. The survey focused on detecting emission and absorption due to transitions between the $J=0$ and $J=1$ rotational states of CO. Throughout the paper we write this with the notation of `CO(1-0)' when making general reference the line and when discussing its emission. We also use `CO(0-1)' specifically in reference to its absorption. This line acts as a tracer for molecular hydrogen at temperatures of up to a few tens of Kelvin; H$_{2}$ is significantly more abundant, but not directly observable at these low temperatures\footnote{Assuming a carbon abundance equal to that of the Milky Way gas phase, and that all gas phase carbon exists in CO molecules, the ratio of carbon monoxide to molecular hydrogen is \mbox{CO/H$_{2}=3.2\times 10^{-4}$ \citep{Sofia}.}}. 
As well as the spectral window in which CO lines were anticipated, the brightest cluster galaxies were observed in neighbouring spectral windows in order to estimate the flux densities of their continuum sources. These observations, which are done at a much lower spectral resolution, also provide serendipitous detections of CN lines from the N = 0 - 1 transition. CN molecules are primarily produced by photodiscociation reactions of HCN, and its presence is therefore indicative of dense, molecular gas in the presence of a strong ultraviolet radiation field \citep[for a detailed overview of the origins of CN, see][]{Boger2005}. Additionally, models have shown that CN production at high column densities can be induced by strong X-ray radiation near active galactic nuclei \citep{Meijerink2007}. 

As well as CO and CN lines, in one case SiO absorption is also detected. This dense gas tracer is often indicative of shocks due to outflows and jet-cloud interactions, and its abundance is highest around galactic centres \citep{Rodriguez2006, Rodriguez2010}.

In Table \ref{tab:observations_detections_table}, we summarise the observations and detections of CO, CN and H\thinspace\small I\normalsize\space lines which have been made both in this survey and archival observations. The top section of the table gives details for sources in which we see some form of molecular absorption, discussed fully in \S\ref{sec:Detections}. The middle section shows sources later discussed in \S\ref{sec:emitting_sources} in which we see emission, but no absorption. The lower section gives details for sources in which we see no molecular absorption or emission, discussed in \S\ref{sec:NonDetections}. This table provides a useful reference for the reader throughout the paper and helps to place our detections within a wider context. Details of the observations for all sources in which we find evidence of molecular gas from emission and/or absorption lines are given in Table \ref{tab:observations_table}. 

Below we provide a short description of the previous observations of each galaxy in our survey. We also highlight any previous detections of H\thinspace\small I\normalsize\space absorption, a tracer of warm atomic gas. In ambiguous cases where a source's name is often used to describe both the individual brightest cluster galaxy and the wider cluster, we use the name in reference to the former.

\begin{itemize}
    \item \textbf{Hydra-A} is a giant elliptical galaxy with a close to edge-on disk of dust and molecular gas lying at the centre of an X-ray luminous cluster \citep{Hamer2014}. Hydra-A is an archetype of a brightest cluster galaxy lying in a cooling flow, with powerful radio jets and lobes projected outwards from its centre \citep{Taylor1990}. These are surrounded by cavities in the X-ray emitting gas of the intracluster medium created by repeated AGN outbursts \citep{McNamara2000, Wise2006}. Previous observations of Hydra-A show extremely strong CO(1-2) absorption against the bright radio core ($\tau_{\textnormal{max}} \sim 0.9$) due to molecular gas moving away from the galaxy centre at a few tens of \kms \citep{Rose2019}. H\thinspace\small I\normalsize\space absorption has been detected against the core of the galaxy with a peak optical depth of $\tau$~=~0.0015 \citep{Taylor1996}.
    \item \textbf{S555} is a relatively anonymous low X-ray luminosity cluster selected by the REFLEX survey \citep{Bohringer2004} which has a strong compact
    radio source \citep{Hogan2015}, is known to be core dominated and has a significant radio and gamma-ray flux density \citep{Dutson2013}. Against the core of the galaxy, H\thinspace\small I\normalsize\space absorption has been searched for, providing an upper limit of $\tau_{\textnormal{max}}<0.013$ \citep{Hogan_thesis}.
    \item \textbf{Abell 2390} lies at the centre of a highly X-ray luminous cluster \citep[L$_{X}\sim 10^{45} \textnormal{~erg s}^{-1}$,][]{Ebeling1996} with a significant cooling flow of 300 M$_{\odot}$ yr$^{-1}$ \citep{Allen2001}. The galaxy has extended optical emission lines \citep{LeBorgne1991} and contains a significant mass of dust showing up as strong absorption in optical and submillimetre continuum emission \citep{Edge1999}. Against the core of the galaxy, H\thinspace\small I\normalsize\space absorption has been detected with $\tau_{\textnormal{max}}=0.084 \pm 0.011$ \citep{Hogan_thesis}.
    \item \textbf{RXCJ0439.0+0520} has been found to be highly variable in the radio, with significant changes occurring in its 15 GHz spectrum over year long timescales \citep{Hogan2015}. Optical emission line studies also show a significant H$\alpha$ luminosity of \mbox{$6 \times 10^{40}$ erg s$^{-1}$} \citep{Hamer2016}. Against the core of the galaxy, H\thinspace\small I\normalsize\space absorption has been searched for, providing an upper limit of $\tau_{\textnormal{max}}<0.133$ \citep{Hogan_thesis}.
    \item \textbf{Abell 1644} is a poorly studied source lying at the centre of the brighter of two X-ray peaks in its host cluster, which itself has evidence of gas sloshing \citep{Johnson2010}. H\thinspace\small I\normalsize\space absorption has been detected, though is yet to be published. 
    \item \textbf{NGC 5044} is a highly perturbed brightest cluster galaxy which contains a significant mass of multiphase gas. It is surrounded by numerous cavities and X-ray filaments which have been inflated by the AGN \citep{Buote2004,David2011,Gastaldello2013}. CO(2-1) observations show significant emission and give an inferred molecular gas mass of 6.1$\times 10^{7}~\textnormal{M}_{\odot}$ \citep{Temi2018}. CO(1-2) absorption has also been observed due to a series of molecular gas clouds lying along the line of sight to the continuum source, with velocities of approximately \mbox{$250$ \kms} \citep{David2014}.
    \item \textbf{NGC 6868} is poorly studied, though it has been found to have a flat spectrum with a core flux density of 105 mJy at 5 GHz \citep{Hogan2015b}. H\thinspace\small I\normalsize\space absorption has been observed against the galaxy's core at a velocity of v $\sim 50$ \kms\space and FWHM $\sim 80$ \kms\space (Tom Oosterloo, private communications)
    \item \textbf{Abell 2597} is a giant elliptical brightest cluster galaxy surrounded by a dense halo of hot, X-ray bright plasma of megaparsec scales. Observations by \citet{Tremblay2016, Tremblay2018} show CO(2-1) emission at the systemic velocity of the galaxy. There are also three distinct regions of CO(1-2) absorbing molecular gas along the line of sight to the galaxy's radio core, with optical depths of $\tau \sim 0.2-0.3$ and velocities of $240-335$ \kms. 
    \item \textbf{RXCJ1350.3+0940} lies in an extremely strong cool-core cluster which, while selected as part of the ROSAT Bright Source catalogue \citep[RBS1322,][]{Schwope2000}, was  misidentified as a BL-LAC \citep{Massaro2009, Richards2011, Green2017} because it is dominated by a 300~mJy, flat-spectrum radio core. Despite having radio, optical, MIR and sub-mm properties which are similar to many of the best known cool-core clusters (e.g. Abell 1068, Abell 1835 and Zw3146), overall the galaxy remains poorly studied \citep{Hogan2015b,Green2016}. However, around the core of the galaxy, H\thinspace\small I\normalsize\space absorption has been searched for, giving an upper limit of $\tau_{\textnormal{max}}< 0.0054$ \citep{Hogan_thesis}.
    \item \textbf{MACSJ1931.8-2634} lies within an extremely X-ray luminous cool-core containing large cavities and an equivalent mass cooling rate of $\sim 700$ $\textnormal{M}_{\odot} \textnormal{ yr}^{-1}$ in the central 50 kpc \citep{Allen2004, Allen2008}. Clear structure exists within the cluster core and the brightest cluster galaxy itself is strongly elongated in the North-South direction \citep{Ehlert2011}. ALMA data at higher frequencies have recently been published by \citep{Fogarty19} but no attempt to determine the extent of any absorption against the core was made in that paper.
    \item \textbf{RXCJ1603.6+1553} is another relatively poorly studied cluster, likely due to its brightest cluster galaxy being dominated by a flat-spectrum radio core. Like RXCJ1350.3+0940, the source was selected in the ROSAT Bright Source catalogue (RBS1552) but the bright radio core led to the X-ray source being classified as a BL~Lac. HI absorption has been detected close to the galaxy's systemic recession velocity with a peak optical depth of $\tau_{\textnormal{max}}=0.125$ and FWHM $\sim$~400~\kms \citep{Gereb2015}. 
    \item \textbf{MACSJ0242.5-2132} contains one of the most radio powerful core sources in the sample presented in \citet{Hogan2015b}.  The redshift of this source at $z=0.31$ means that the H\thinspace\small I\normalsize\space absorption is strongly affected by RFI, so no sensitive observations of this source have yet been attempted. 
    \item \textbf{Abell 3112} has a strong source at its core in our ALMA continuum image consistent with the position of the published Long Baseline Array observation. However, a second unresolved source is visible to the North-West of the core consistent with a compact, off nuclear source seen in archival HST imaging. The galaxy has an upper limit for H\thinspace\small I\normalsize\space absorption of \mbox{$\tau_{\textnormal{max}} <0.007$}, made with the Australia Telescope Compact Array (ATCA) and shown in \citet{Hogan_thesis}.
    \item \textbf{Abell 496} is poorly studied, though has an upper limit for H\thinspace\small I\normalsize\space absorption from the Very Large Array (VLA) presented by \citet{Hogan_thesis}.
    \item \textbf{RXCJ0132.6-0804} is highly X-ray luminous \citep[3.6$\times10^{44}$ erg s$^{-1}$][]{Bohringer2002} and core dominated, with evidence of AGN activity \citep{Hamer_thesis}. It also has a highly variable radio flux density, with up to $\sim 80$ per cent variability at 150 GHz found by \citet{Hogan2015}. 
    \item \textbf{Abell 2415} is poorly studied, though has an as yet unpublished H\thinspace\small I\normalsize\space absorption detection from the {\it Jansky} VLA from 2015 (PI: Edge) with an estimated peak optical depth of \mbox{$\tau_{\textnormal{max}} =0.02$}. 
    \item \textbf{Abell 3581} hosts one of the best studied, low redshift and radio loud brightest cluster galaxies, PKS~1404-267 \citep{Johnstone1998,Johnstone2005}. The cluster shows evidence of multiple AGN outbursts \citep{Canning2013} and ALMA observations detect strong CO(2-1) emitting gas filaments \citep{Olivares2019}. \cite{Johnstone1998} present a VLA spectrum showing no significant H\thinspace\small I\normalsize\space absorption.
    \item \textbf{RXCJ1356.0-3421} has X-ray properties consistent with a strong cooling flow. It should therefore have been included in the REFLEX cluster sample that is one of the two primary X-ray samples that make up the parent sample for this study, but was assumed to be AGN dominated \citep{Somboonpanyakul2018}. H\thinspace\small I\normalsize\space absorption with \mbox{$\tau_{\textnormal{max}}=0.125$} and a full-width-zero-intensity of \mbox{$\sim$ 500 \kms} has been published by \citep{Veron-Cety2000}, implying that a significant column density of atomic gas is present in this system. 
\end{itemize}

\section{Data processing and the origin of the CN-A and CN-B absorption lines}
\label{sec:DataReduction}

\begin{figure*}
	\includegraphics[width=\textwidth]{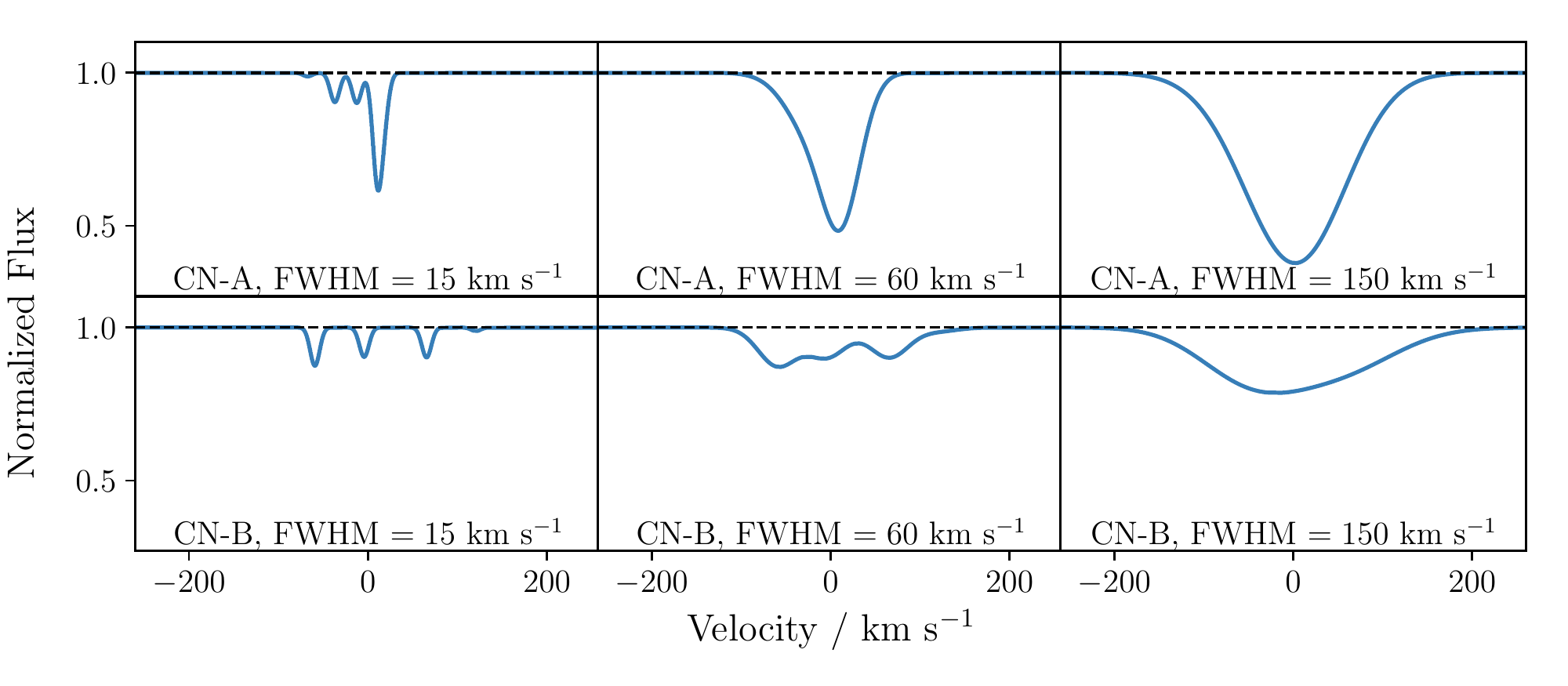}
     \caption{Our observations show detections of two CN lines, labelled as CN-A and CN-B throughout the paper. These are respectively formed by the combination of five and four hyperfine structure lines, details of which are shown in Table \ref{tab:CN_rotational_lines}. Our low spectral resolution CN observations do not resolve this hyperfine structure, and as such, we treat each of the two groups of lines as single Gaussians during our analysis. Here, we simulate the appearance of these two sets of lines for increasing FWHM and a constant, arbitrary peak intensity to show how they appear as they blend together. The velocities are calculated using the intensity weighted mean of the CN-A and CN-B line centres rather than the individual line frequencies (see Table \ref{tab:CN_rotational_lines}). We also apply the same calculations to our CN spectra. Even without including noise, the hyperfine structure lines merge together as the FWHM increases towards the CN channel widths of the observations shown in Fig. \ref{fig:CO_10_Absorbers_first_plot} and \ref{fig:CO_10_Absorbers_second_plot} ($\sim$ 60 km s$^{-1}$). Gaussian fits to these lines show that CN-B has the larger FWHM, consistent with the observations. The slight asymmetry which is particularly prominent in the blended CN-B line is also seen in the majority of the spectra shown in \mbox{Fig. \ref{fig:CO_10_Absorbers_first_plot} and \ref{fig:CO_10_Absorbers_second_plot}}.}
    \label{fig:CN_Line_Simulation_Plot}
\end{figure*}

\begin{table*}
	\centering
	\begin{tabular}{ccc} 

		\hline
		Line & Rest frequency (GHz) \\
		\hline
		CO(1-0) & 115.271208 & \\
		CO(2-1) & 230.538000 & \\
		CN-A & 113.49485 & \\
		CN-B & 113.16883 & \\
		SiO(3-2) & 130.268610 & \\
		\hline
		\\ \\
		\hline
		CN transition & Rest frequency & Relative \\
		$N,J,F \rightarrow N^{\prime},J^{\prime},F^{\prime}$ & (GHz) & intensity\\
		\hline
         1,3/2,3/2 $\rightarrow$ 0,1/2,1/2 & 113.48812 & 0.125 \\
         1,3/2,5/2 $\rightarrow$ 0,1/2,3/2 & 113.49097 & 0.333 \\
         1,3/2,1/2 $\rightarrow$ 0,1/2,1/2 & 113.49964 & 0.099 \\
         1,3/2,3/2 $\rightarrow$ 0,1/2,3/2 & 113.50890 & 0.096 \\
         1,3/2,1/2 $\rightarrow$ 0,1/2,3/2 & 113.52043 & 0.012 \\
         \hline
         1,1/2,1/2 $\rightarrow$ 0,1/2,1/2 & 113.12337 & 0.012 \\
		 1,1/2,1/2 $\rightarrow$ 0,1/2,3/2 & 113.14415 & 0.098 \\
		 1,1/2,3/2 $\rightarrow$ 0,1/2,1/2 & 113.17049 & 0.096 \\
         1,1/2,3/2 $\rightarrow$ 0,1/2,3/2 & 113.19127 & 0.125 \\
		\hline 
		
	\end{tabular}
    \caption{\textbf{Top:} The rest frequencies used throughout this paper when converting from frequencies to velocities. The frequencies of the CN-A and CN-B lines are the intensity weighted means of the component hyperfine structure lines. \textbf{Bottom:} The individual hyperfine structure lines of CN existing in the frequency range of the observed absorption regions \citep{Splatalogue_CDMS}. From the hyperfine structure lines listed here, the first five and last four each blend together to form two separate lines in the relatively low resolution observations. When fitting to these two lines, we use an intensity weighted mean for the line frequency. For our lines labelled \mbox{CN-A}, this frequency is \mbox{113.49485 GHz} and for those labelled \mbox{CN-B}, is \mbox{113.16883 GHz}. In Fig. \ref{fig:CN_Line_Simulation_Plot}, we demonstrate how this hyperfine structure of CN blends to give the CN-A and CN-B lines.}
    \label{tab:CN_rotational_lines}
\end{table*}

The data presented in this paper were handled using CASA version 5.1.1, a software package which is produced and maintained by the National Radio Astronomy Observatory (NRAO) \citep{CASA}. The calibrated data were produced by the ALMA observatory and following their delivery, where needed we made channel maps at maximal spectral resolution. The self-calibration and continuum subtraction of the images were done as part of the pipeline calibration. When converting the frequencies of the observed CO(1-0) spectra to velocities, we use a rest frequency of \mbox{$f_{\textnormal{CO(1-0)}} = 115.271208$ GHz}. The CN absorption is more complex than that of CO due to its hyperfine structure and the lower spectral resolution with which it was observed. Two lines are seen in CN for each absorption region detected, with relative peak line strengths of approximately 2:1. These two poorly resolved absorption features are themselves composed of a mixture of hyperfine structure lines, details of which are given in Table \ref{tab:CN_rotational_lines} and Fig. \ref{fig:CN_Line_Simulation_Plot}. The CN lines covered by our observations are of the N = 0 - 1 transition, which consists of nine hyperfine structure lines split into two distinct groups. The stronger group being J = 3/2 - 1/2 transitions and the weaker group J = 1/2 - 1/2  transitions. Throughout the paper, the label CN-A is used to denote the stronger absorption line, and CN-B to denote the weaker line. We use the intensity weighted mean of the component hyperfine structure lines to calculate the rest frequencies, resulting in \mbox{$f_{\textnormal{CN-A}} = 113.49485$ GHz} and \mbox{$f_{\textnormal{CN-B}} = 113.16883$ GHz}. For the single detection of \mbox{SiO(2-3)}, we use a rest frequency of \mbox{$f_{\textnormal{SiO(3-2)}} = 130.268610$ GHz}.

We use a range of sources to determine the velocity of the emission and absorption features in each galaxy relative to its recession velocity. The velocities we use for each galaxy and their sources are listed in Table \ref{tab:recession_velocities_table}.

\begin{table*}
\caption{Stellar redshifts and their corresponding velocities used as zero-points for the spectra shown in Fig. \ref{fig:CO_10_Absorbers_first_plot}, \ref{fig:CO_10_Absorbers_second_plot} and \ref{fig:CO_10_Absorbers_third_plot}. All redshifts are barycentric and use the optical convention. The stellar redshifts of Hydra-A, Abell 1644 and NGC 5044 are taken from Multi Unit Spectroscopic Explorer (MUSE) observations (ID: 094.A-0859). Further details of the MUSE stellar redshift used for Abell 2597 can be found in \citet{Tremblay2018}. The stellar redshifts of RXCJ1350.3+0940 and RXCJ1603.6+1553 are from the Sloan Digital Sky Survey (SDSS) \citep{SLOAN}. The stellar redshift of MACS1931.8-2634 is taken from \citet{Fogarty19} and is found using MUSE observations.
Crosschecking with FOcal Reducer and low dispersion Spectrograph (FORS) observations of S555, Abell 1644, NGC 5044, and Abell 2597 provides redshifts in good agreement with those listed below. The redshifts used for Abell 2390 and RXCJ0439.0+0520 are taken from Visible Multi-Object Spectrograph (VIMOS) observations previously presented by \citet{Hamer2016} and are based primarily on stellar emission lines. The observed wavelengths of the single stellar absorption line in these two VIMOS spectra are consistent with the quoted redshifts. The VIMOS redshift of RXCJ0439.0+0520 also matches that derived from the multiple absorption lines found from an archival William Hershel Telescope (WHT) observation using the ISIS spectrograph.}
	\centering
	\begin{tabular}{lccr}
	
\hline
Source & Redshift & Recession velocity (km s$^{-1}$) & Redshift source \\
\hline

  Hydra-A & 0.0544$\pm$0.0001 & 16294$\pm$30 & MUSE \\
  S555 & 0.0446$\pm$0.0001 & 13364$\pm$30 & MUSE \\
  Abell 2390 & 0.2304$\pm$0.0001 & 69074$\pm$30 & VIMOS \\
  RXCJ0439.0+0520 & 0.2076$\pm$0.0001 & 62237$\pm$30 & VIMOS \\
  Abell 1644 & 0.0473$\pm$0.0001 & 14191$\pm$30 & MUSE \\
  NGC 5044 & 0.0092$\pm$0.0001 & 2761$\pm$30 & MUSE \\
  NGC 6868 & 0.0095$\pm$0.0001 & 2830$\pm$30 & FORS \\
  Abell 2597 & 0.0821$\pm$0.0001 & 24613$\pm$30 & MUSE \\
  RXCJ1350.3+0940 & 0.13255$\pm$0.00003 & 39737$\pm$10 & SDSS \\
  MACS1931.8-2634 & 0.35248$\pm$0.00004 & 105670$\pm$10 & MUSE \\
  RXCJ1603.6+1553 & 0.10976$\pm$0.00001 & 32905$\pm$3 & SDSS \\
		\hline 
	\end{tabular}
    \label{tab:recession_velocities_table}
\end{table*}

\begin{figure*}
	\includegraphics[width=\textwidth]{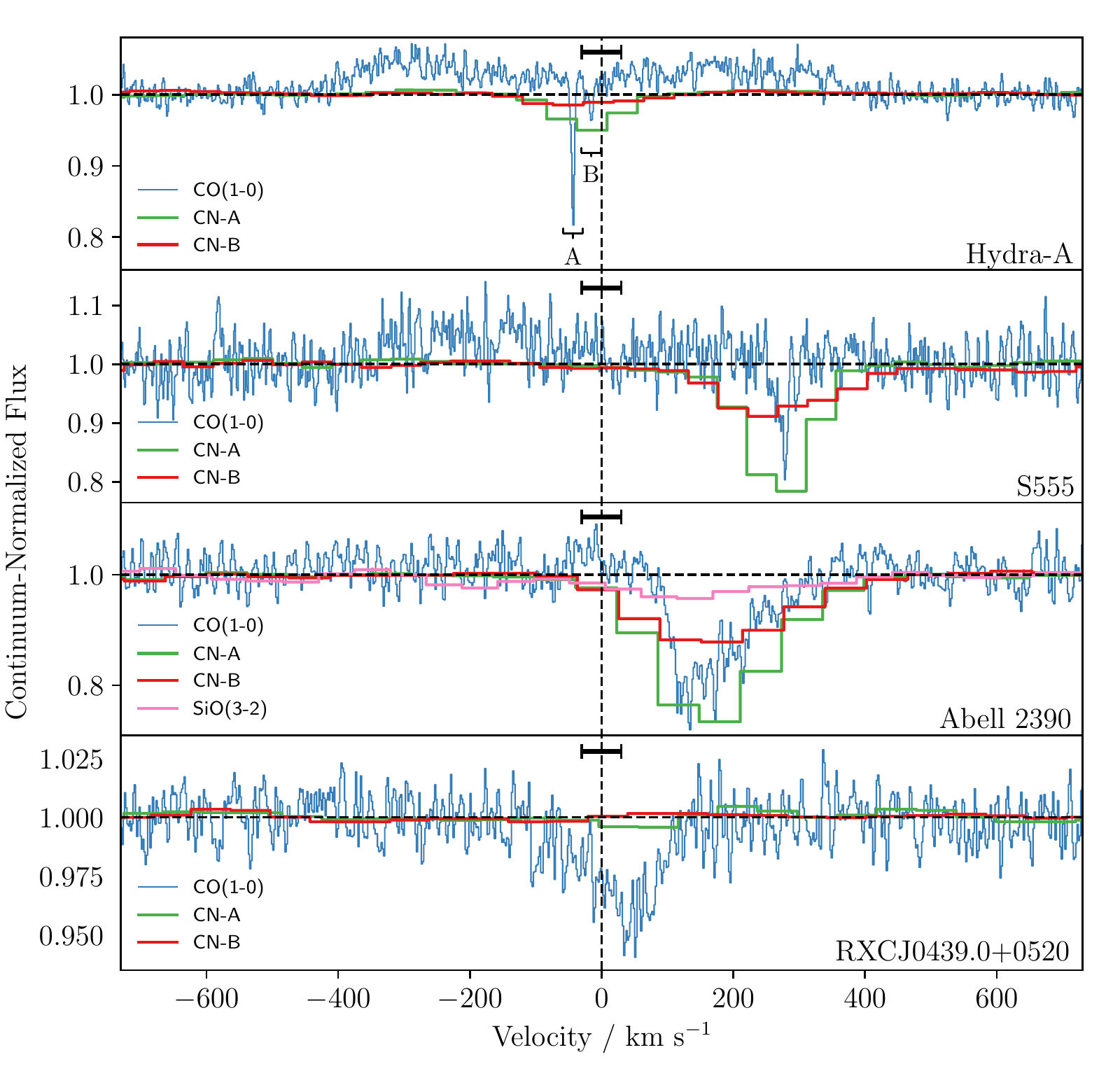}
    \caption{CO(1-0) and CN-A/CN-B spectra from along the line of sight to each object's continuum source, extracted from a region with a size equal to the synthesized beam's FHWM. The spectra shown here are those with $\geq 3 \sigma$ detections of CO and/or CN absorption out of the sample of 18 observed. Each of the two CN lines shown is produced by the combination of several of the molecule's hyperfine structure lines (see Fig. \ref{fig:CN_Line_Simulation_Plot} and Table \ref{tab:CN_rotational_lines} for further details). One source, Abell 2390, also shows a SiO(2-3) absorption line detection. Where available, we also include archival observations of CO(2-1). The recession velocity on which each spectrum is centred can be found in Table \ref{tab:recession_velocities_table} and the error bars shown in the top-middle of each spectrum indicate the systematic uncertainty of this value. Continued in Fig. \ref{fig:CO_10_Absorbers_second_plot}.}
    \label{fig:CO_10_Absorbers_first_plot}
\end{figure*}
\begin{figure*}
    \includegraphics[width=\textwidth]{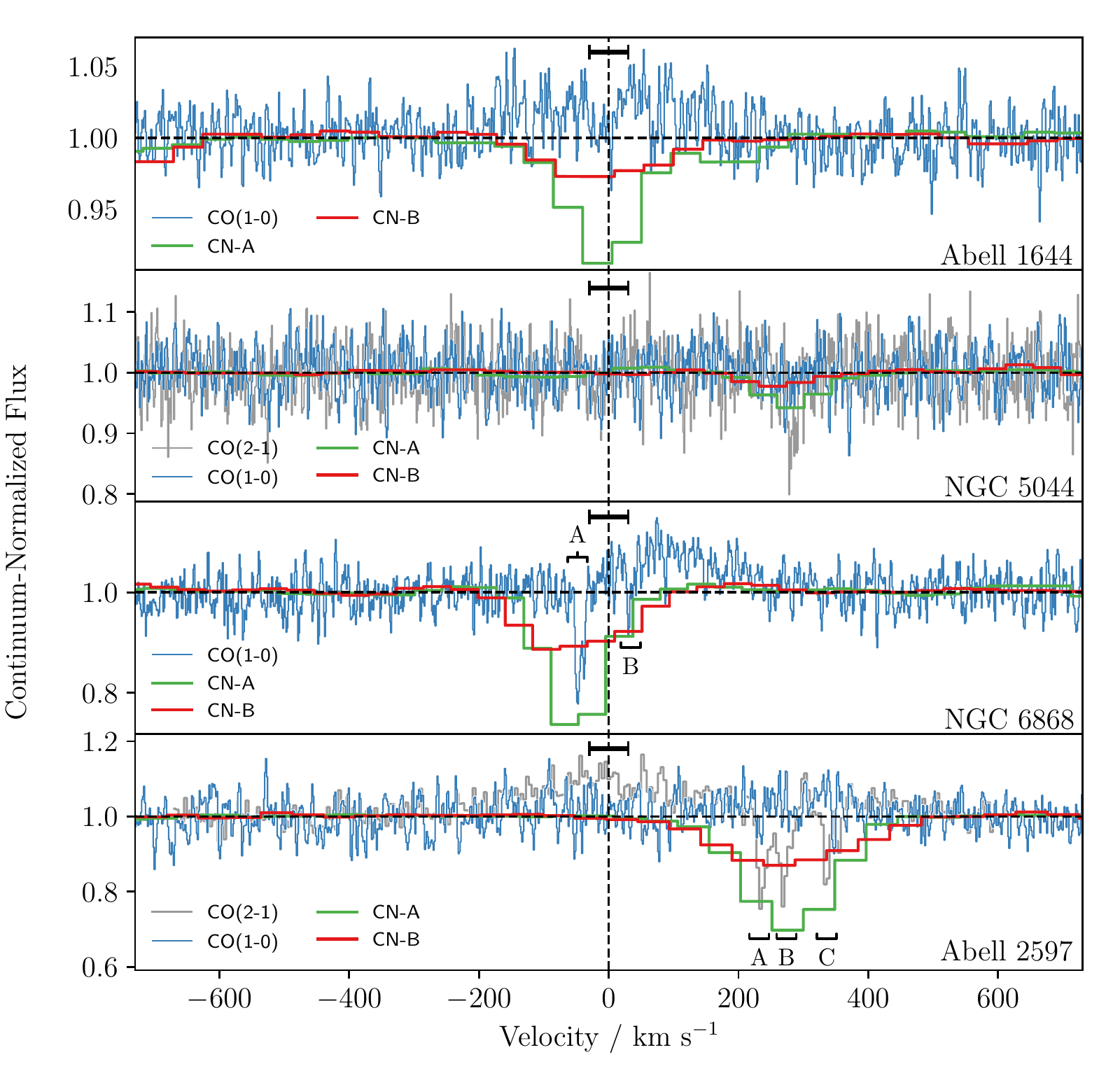}
    \caption{CO(1-0) and CN-A/CN-B spectra extracted from regions centred on each object's continuum source and with a size equal to the synthesized beam's FHWM. Continued from Fig. \ref{fig:CO_10_Absorbers_first_plot}.}
    \label{fig:CO_10_Absorbers_second_plot}
\end{figure*}

\section{Molecular absorption in the cores of eight brightest cluster galaxies}
\label{sec:Detections}

From the sample of 18 brightest cluster galaxies observed, we find $\geq 3 \sigma $ evidence of molecular absorption in eight. Their absorption spectra, each extracted from a region centred on the continuum source with a size equal to the synthesized beam's FHWM, are shown in Fig. \ref{fig:CO_10_Absorbers_first_plot} and \ref{fig:CO_10_Absorbers_second_plot}. The continuum emission against which we see this absorption is unresolved in all of these sources, and therefore the absorption itself is not spatially resolved.

In Table \ref{tab:results_table}, we show the central velocity, FWHM, amplitude, peak optical depth and velocity integrated optical depth of the emission and absorption features. The values and errors are calculated by performing Monte Carlo simulations which re-simulate the noise seen in each spectrum, along the same lines as described in \citet{Rose2019}. To summarise, for each observed spectrum the noise level is estimated from the root mean square (rms) of the continuum source's emission. This is calculated after excluding the region where the emission is clearly visible. Following this, \mbox{10 000} simulated spectra are produced. To make each simulated spectrum, a Gaussian distribution is created for each velocity channel. This distribution is centred at the intensity in the observed spectrum for that particular velocity channel, with a variance equal to the rms noise squared. A value for the intensity is drawn at random from the Gaussian distribution. When this has been done for all velocity channels, a simulated spectrum is produced. For each of the \mbox{10 000} simulated spectra, Gaussian lines are fitted to the absorption and emission line features. The values which delimit the 15.865 per cent highest and lowest results for each of the fits give the upper and lower ${1\sigma}$ errors, meaning that 68.27 per cent of the fitted parameters will lie within the ${1\sigma}$ range.

Below we describe the emission and absorption features seen in each source.

\begin{itemize}
    \item \textbf{Hydra-A} shows double peaked CO(1-0) emission due to the edge-on orientation of its disk and the large beam size of the observations. Close to the zero velocity point, two \mbox{CO(0-1)} absorption features can be seen, one of which is strong and extremely narrow ($\tau_{\textnormal{max}}=0.22^{+0.1}_{-0.1}$, FWHM $= 5.2^{+0.4}_{-0.3}$ \kms). These are also matched by CN-A/CN-B absorption lines. This feature appears stronger still in previous CO(1-2) absorption, where the optical depth peaks at $\tau_{\textnormal{max}}= 0.9$. In order to show the CO(0-1) and CN-A/CN-B absorption more clearly, we do not show the CO(1-2) absorption line of Hydra-A due to its significantly larger optical depth. It can however be found in \citet{Rose2019}. 
    \item \textbf{S555} shows a \mbox{CO(1-0)} emission line, as well as \mbox{CO(0-1)} and CN-A/CN-B absorption lines at large redshifted velocities of $\sim 270$ \kms. These high velocities imply significant line of sight motion towards the mm-continuum source. The combined integrated optical depth of the CN-A/CN-B absorption lines is around 20 times larger than that of CO(0-1), implying a low molecular ratio of CO/CN. 
    \item \textbf{Abell 2390} has no visible CO(1-0) emission, but does show CO(0-1), CN-A/CN-B and SiO(2-3) absorption lines. All of these lines are wide, slightly skewed Gaussians centred at a velocity of $\sim 170$ \kms. Despite its large FWHM, the absorption feature has a sharp onset in the high spectral resolution CO(1-0) spectrum.
    \item \textbf{J0439+05} has no CO(1-0) emission, though a wide CO(0-1) absorption feature (FWHM$ = 126^{+10}_{-10}$ \kms) is present close to the zero velocity point, which is unique among the sample in that there are no corresponding CN-A/CN-B lines.
    \item \textbf{Abell 1644} has a broad CO(1-0) emission region, but no statistically significant CO(0-1) absorption. However, strong CN-A/CN-B absorption is present at the centre of the CO(1-0) emission.
    \item \textbf{NGC 5044}, which was previously found to have redshifted CO(1-2) absorption at $\sim 300$ \kms\space\citep{David2014}, has corresponding CN-A/CN-B lines with a total of around four times the velocity integrated optical depth. However, perhaps due to the realtively high noise level, there is no statistically significant CO(0-1) absorption feature. Likewise, there is no clear CO(1-0) emission.
    \item \textbf{NGC 6868} has the narrowest CO(1-0) emission feature of the sample (FWHM = 207$^{+18}_{-18}$ \kms), though it is consistent with the range of line widths found in single dish studies \citep{Edge2001,SalomeCombes2003}. At the blueshifted edge of this emission there are two narrow CO(0-1) absorption features. CN absorption centred on the stronger, more blueshifted of the two CO(0-1) absorption features is also present. As with the other sources, its CN-A/CN-B absorption has a much larger velocity integrated optical depth than that of the CO(0-1). By this measure, the two CN absorption lines are around 10 times stronger than those of CO(0-1).
    \item \textbf{Abell 2597} has previously been shown to have \mbox{CO(2-1)} emission whose central velocity matches that of the galaxy's stellar recession velocity. There are also three narrow absorption features at velocities of between 240 and 335 \kms\space\citep{Tremblay2016}. These absorption features are also detected at low resolution in CN-A/CN-B, but not in CO(0-1). A weak CO(1-0) emission line is present in the spectrum. However, this is centred at approximately the same velocity as the CO(1-2) and CN-A/CN-B absorption features, rather than close to the systemic velocity where the CO(2-1) emission is seen. This velocity difference between the weak but broad CO(1-0) emission and stronger CO(2-1) emission indicates that the warmer gas, which likely lies closer to the core, traces gas with different dynamics compared with the colder gas traced by the CO(1-0). 
\end{itemize}

In many cases, it should be noted that our calculations of the optical depths are simply lower limits. This is due to the difficulty of establishing to what extent emission is compensating for absorption in some spectra. For example, in Abell 2390 there are hints of emission either side of the absorption region, which could reduce the level of absorption we infer. Where the emission is clearer, such as in NGC 6868 and Abell 1644\footnote{In the case of Abell 1644, the tentative absorption feature at $v\sim 0$ \kms\space is of less than 3$\sigma$ significance.}, we can compensate for it. This is done by fitting and subtracting a Gaussian line to the CO(1-0) emission after excluding the velocity channels in which the absorption regions lie. For the CN-A/CN-B lines, this effect is unlikely to have an impact because it is only expected to be present very weakly in emission \citep{Wilson2018}.

\section{Sources with emission which lack absorption lines}
\label{sec:emitting_sources}

\begin{figure*}
	\includegraphics[width=\textwidth]{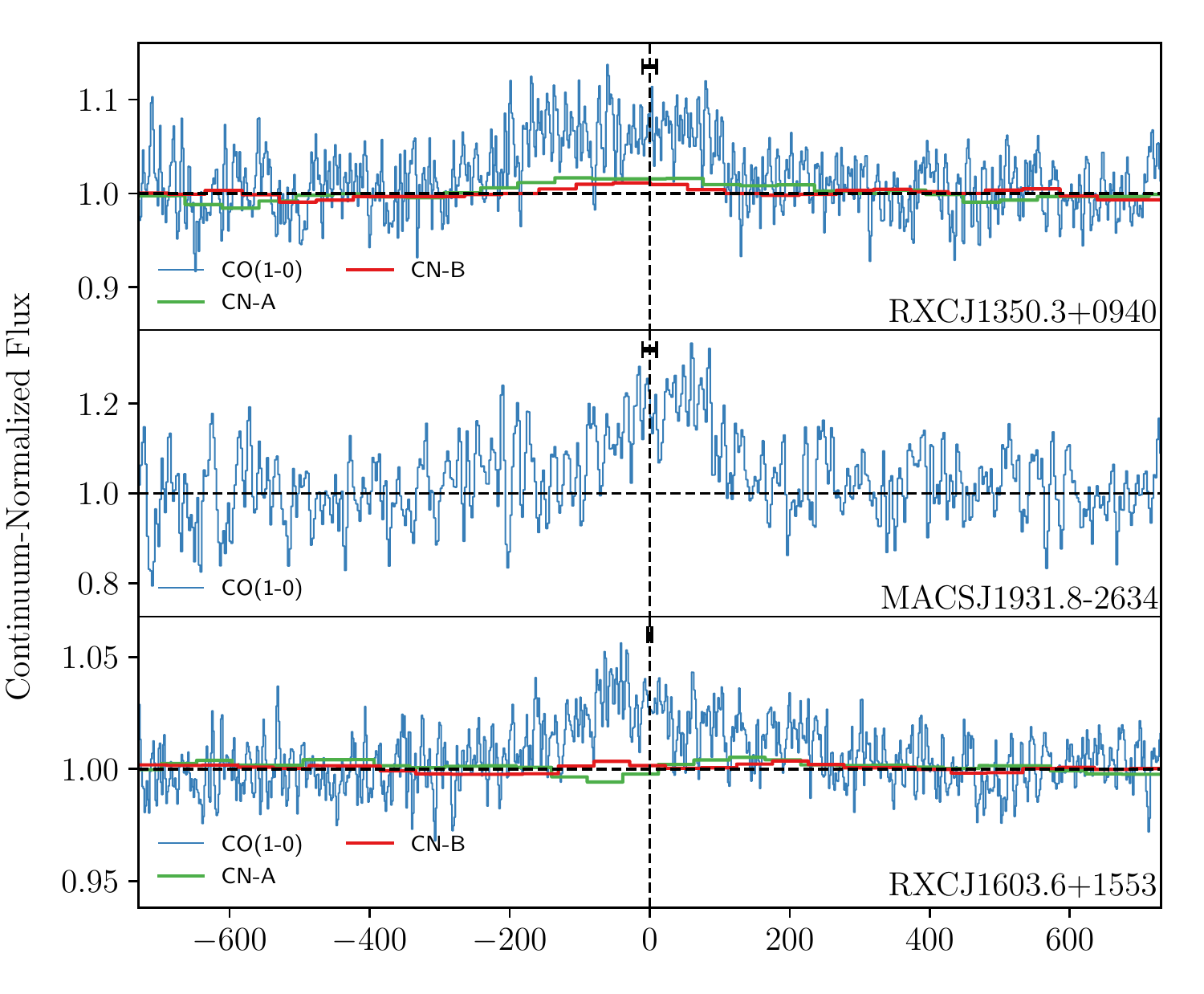}
    \caption{CO(1-0) and CN-A/CN-B spectra of sources which do not have $\geq 3 \sigma$ absorption features of either CO(0-1) or CN-A/CN-B despite having significant masses of molecular gas in their cores, evidenced by clear CO(1-0) emission. These spectra are each extracted from a region centred on the continuum source with a size equal to the synthesized beam's FHWM. This is the smallest region from which they can feasibly be extracted from and maximises the strength of any tentative absorption features in the spectra. The error bars shown in the top-middle of each spectrum indicate the systematic uncertainty in the recession velocity on which each spectrum in centered.}
    \label{fig:CO_10_Absorbers_third_plot}
\end{figure*}

As well as the eight brightest cluster galaxies which have $\geq 3 \sigma$ evidence of CO(0-1) and/or CN-A/CN-B absorption lines there are three systems which have clear emission, but no absorption features. The spectra of these sources are shown in Fig. \ref{fig:CO_10_Absorbers_third_plot} and the best fit parameters for the absorption features are given in the lower section of Table \ref{tab:results_table}. These spectra are once again extracted from a region which is centered on each object's continuum source and with a size equal to the synthesized beam's FHWM. This is the smallest region from which the spectra can feasibly be extracted and it therefore maximises the strength of any tentative absorption features which may be present.

The three sources which show CO(1-0) emission but lack any absorption features are:

\begin{itemize}
    \item \textbf{RXCJ1350.3+0940}, which also shows clear emission from the CN-A and CN-B lines.
    \item \textbf{MACSJ1931.8-2634}, which is also known to show extended CO(3-2) and CO(4-3) emission \citep{Fogarty19}.
    \item \textbf{RXCJ1603.6+1553}, in which H\thinspace\small I\normalsize\space absorption has been detected close to the systemic recession velocity of the galaxy with a peak optical depth of $\sim 10$ and FWHM = $\sim 400$ \kms.
\end{itemize}

\section{Sources without emission and absorption lines}
\label{sec:NonDetections}

As well as the detections shown in Fig. \ref{fig:CO_10_Absorbers_first_plot}, \ref{fig:CO_10_Absorbers_second_plot} and \ref{fig:CO_10_Absorbers_third_plot}, a large number of the sources observed in our ALMA survey show no clear evidence of emission, or of absorption along the line of sight to their continuum source. The details of their observations are given in Table \ref{tab:observations_table_non_detections} of Appendix \ref{sec:appendix,nondetections}. We do not show the spectra of these sources, though their observations, including all continuum images, are publicly available via the ALMA Science Archive from September 20 2019.

In total there are seven sources observed for which we see no  $\geq 3 \sigma$ evidence of molecular gas along the line of sight to their bright radio cores from emission or absorption, all of which are listed below.
\begin{itemize}
    \item \textbf{MACSJ0242.5-2132}
    \item \textbf{Abell 3112}
    \item \textbf{Abell 496}
    \item \textbf{RXCJ0132.6-0804}
    \item \textbf{Abell 2415}
    \item \textbf{Abell 3581}
    \item \textbf{RXCJ1356.0-3421}
\end{itemize}

Additionally, none of the galaxies listed above have CO(1-0) emission which is visible on larger galaxy-wide scales, with the exception of RXCJ0132.6-0804. The extended CO(1-0) emission seen in this system follows the morphology previously found with optical emission lines \citep{Hamer2016}.

In systems such as those observed in our ALMA survey, the line of sight covering fraction of molecular gas is expected to be less than its maximum physical value of 1. In other words, molecular gas is not expected to exist along all lines of sight to the galaxies' bright radio cores. Therefore, the lack of absorption lines in the systems listed above does not necessarily mean that significant masses of cold molecular gas are absent. Overall, the eight absorbing systems we find from the sample of 18 observed implies a line of sight covering fraction in line with expectations and is similar to that predicted by accretion simulations, such as those by \citet{Gaspari2018}.

However, sources which have both CO(1-0) and CO(2-1) observations (Hydra-A, NGC 5044 and Abell 2597) all show the higher energy CO(2-1) line to be significantly stronger in both emission and absorption. It is therefore likely that our 8/18 detection rate is only indicating the covering fraction of particularly cool molecular gas at a up to few tens of Kelvin. Above $\sim 50$ K, the fraction of CO molecules occupying the ground state energy level is negligible, and so CO(0-1) absorption from this line is no longer seen. Large proportions of the molecular gas in the cores of these galaxies is likely to exist at higher temperatures not traced well by CO(0-1), as shown by Hydra-A, NGC 5044 and Abell 2597. Therefore, the total covering fraction of molecular gas is likely to be higher than indicated by the CO(1-0) and CN-A/CN-B\footnote{Although our CN observations appear $\sim$ 10 times stronger than those of CO(0-1), they are likely to lack sufficient spectral resolution to reveal all but the widest and strongest absorption lines.} observations alone.

\begin{table*}
	\centering

	\begin{tabular}{lcccccr} 

		\hline
		 Source &  Region & $v_{\text{cen}}$ (\kms) & FWHM (\kms) & Amplitude (mJy) & $\tau_{\text{max}}$ & $\int\tau dv$ (\kms) \\
		\hline
		Hydra-A & CO(1-0) emission & $-275^{+6}_{-7}$& 235$^{+16}_{-16}$ & 3.6$^{+0.1}_{-0.1}$ & - & - \\ \rule{0pt}{3.0ex}  
		 & CO(1-0) emission & 158$_{-13}^{+10}$ & 346$_{-21}^{+26}$ & 2.98$_{-0.09}^{+0.10}$ & - & -\\ \rule{0pt}{3.0ex} 
		 & CO(0-1) absorption `A' & $-43.4^{+0.1}_{-0.1}$ & $5.2_{-0.3}^{+0.4}$ & $-15.9_{-0.8}^{+0.8}$ & $0.22^{+0.01}_{-0.01}$ & $1.17^{+0.06}_{-0.06}$ \\ \rule{0pt}{3.0ex} 
		 & CO(0-1) absorption `B' & -16$_{-1}^{+1}$ & 9$_{-3}^{+5}$ & -4.2$_{-1.0}^{+0.8}$ & $0.05^{+0.02}_{-0.01}$ & 0.5$_{-0.1}^{+0.2}$\\ \rule{0pt}{3.0ex} 
		 & CN-A absorption & -22$_{-2}^{+2}$ & 102$_{-4}^{+4}$ & -4.2$_{-0.1}^{+0.1}$ & 0.052$_{-0.002}^{+0.002}$ & 5.6$_{-0.2}^{+0.2}$ \\ \rule{0pt}{3.0ex} 
		 & CN-B absorption & -32$_{-9}^{+9}$ & 157$_{-15}^{+15}$ & -1.2$_{-0.1}^{+0.1}$ & 0.015$_{-0.001}^{+0.001}$ & 2.5$_{-0.2}^{+0.2}$ \\ 
		 \hline
		S555 & CO(1-0) emission & -186$_{-10}^{+10}$ & 260$_{-18}^{+19}$ & 0.65$_{-0.04}^{+0.04}$ & - & - \\\rule{0pt}{3.0ex} 
		 & CO(0-1) absorption & 276$_{-2}^{+2}$ & 17$_{-6}^{+6}$ & -1.8$_{-0.5}^{+0.3}$ & 0.16$_{-0.03}^{+0.04}$ & $2.7_{-0.5}^{+0.5}$ \\\rule{0pt}{3.0ex} 
		 & CN-A absorption & 270$_{-1}^{+1}$ & 113$_{-3}^{+3}$ & -2.9$_{-0.1}^{+0.1}$ & 0.26$_{-0.01}^{+0.01}$ & 29.6$_{-0.6}^{+0.6}$ \\\rule{0pt}{3.0ex} 
		 & CN-B absorption & 265$_{-4}^{+4}$ & 210$_{-10}^{+11}$ & -1.09$_{-0.04}^{+0.04}$ & 0.089$_{-0.004}^{+0.004}$ & 19.6$_{-0.8}^{+0.8}$ \\
		\hline
		Abell 2390 & CO(0-1) absorption & 164$_{-2}^{+2}$ & 122$_{-4}^{+4}$ & -1.55$_{-0.05}^{+0.05}$ & 0.22$_{-0.1}^{+0.1}$ & 28.2$_{-0.9}^{+0.9}$ \\\rule{0pt}{3.0ex} 
		 & CN-A absorption & 167$_{-1}^{+1}$ & 200$_{-3}^{+3}$ & -2.09$_{-0.02}^{+0.02}$ & 0.31$_{-0.01}^{+0.01}$ & 63.5$_{-0.7}^{+0.8}$ \\\rule{0pt}{3.0ex} 
		 & CN-B absorption & 171$_{-3}^{+3}$ & 251$_{-6}^{+6}$ & -1.00$_{-0.02}^{+0.02}$ & 0.137$_{-0.003}^{+0.003}$ & 36.0$_{-0.7}^{+0.7}$ \\\rule{0pt}{3.0ex} 
		 & SiO(2-3) absorption & 120$_{-30}^{+30}$ & 400$_{-100}^{+100}$ & -0.28$_{-0.05}^{+0.04}$ & 0.037$_{-0.006}^{+0.007}$ & 15$_{-3}^{+3}$ \\
		\hline
		RXCJ0439.0+0520 & CO(0-1) absorption & 35$_{-3}^{+3}$ & 126$_{-10}^{+10}$ & -2.9$_{-0.2}^{+0.2}$ & 0.041$_{-0.002}^{+0.002}$ & 5.4$_{-0.3}^{+0.3}$ \\
		\hline
		Abell 1644 & CO(1-0) emission & 0$^{+12}_{-12}$ & 308$_{-17}^{+19}$ & 0.85$_{-0.6}^{+0.6}$ & - & - \\\rule{0pt}{3.0ex} 
		 & CN-A absorption & -6$_{-1}^{+1}$ & 120$_{-4}^{+4}$ & -3.6$_{-0.1}^{+0.1}$ & 0.089$_{-0.002}^{+0.002}$ & 11.2$_{-0.3}^{+0.3}$ \\\rule{0pt}{3.0ex} 
		 & CN-B absorption & -11$_{-5}^{+5}$ & 170$_{-10}^{+10}$ & -1.09$_{-0.06}^{+0.06}$ & 0.026$_{-0.002}^{+0.002}$ & 4.7$_{-0.3}^{+0.3}$ \\
		\hline
		NGC 5044 & CO(1-2) absorption & 283$_{-1}^{+1}$ & 14$_{-2}^{+2}$ & -2.6$_{-0.4}^{+0.4}$ & 0.14$_{-0.02}^{+0.02}$ & 2.2$_{-0.3}^{+0.4}$ \\\rule{0pt}{3.0ex} 
		 & CN-A absorption & 280$_{-4}^{+4}$ & 101$_{-9}^{+10}$ & -0.85$_{-0.07}^{+0.07}$ & 0.06$_{-0.01}^{+0.01}$ & 6.4$_{-0.5}^{+0.5}$  \\\rule{0pt}{3.0ex} 
		 & CN-B absorption & 258$_{-9}^{+10}$ & 103$_{-17}^{+19}$ & -0.34$_{-0.06}^{+0.06}$ & 0.024$_{-0.004}^{+0.00}$ & 2.6$_{-0.4}^{+0.5}$  \\
		\hline
		NGC 6868 & CO(1-0) emission & 93$^{+7}_{-7}$ & 207$_{-18}^{+18}$ & 1.09 $_{-0.07}^{+0.07}$ &- &- \\\rule{0pt}{3.0ex} 
		& CO(0-1) absorption `A' & -45$_{-1}^{+1}$ & 15$_{-1}^{+2}$ & -3.0$_{-0.2}^{+0.2}$ & 0.24$_{-0.04}^{+0.04}$ & 3.8$_{-0.4}^{+0.4}$ \\\rule{0pt}{3.0ex} 
		& CO(0-1) absorption `B' & 32$_{-2}^{+1}$ & 10$_{-5}^{+5}$ & -1.6$_{-0.4}^{+0.4}$ & 0.12$_{-0.04}^{+0.04}$ & 1.2$_{-0.3}^{+0.3}$ \\\rule{0pt}{3.0ex} 
		 & CN-A absorption & -50$_{-1}^{+1}$ & 101$_{-2}^{+2}$ &  -4.09$_{-0.06}^{+0.06}$ & 0.3$_{-0.01}^{+0.01}$ & 34.4$_{-0.5}^{+0.5}$ \\\rule{0pt}{3.0ex} 
		 & CN-B absorption & -52$_{-2}^{+2}$ & 168$_{-4}^{+4}$ & -1.73$_{-0.04}^{+0.04}$ & 0.12$_{-0.01}^{+0.01}$ & 22.7$_{-0.6}^{+0.6}$ \\
		\hline
		Abell 2597 & CO(1-0) emission & 233$_{-42}^{+46}$ & 400$_{-100}^{+100}$ & 0.24$_{-0.04}^{+0.08}$ & - & - \\\rule{0pt}{3.0ex} 
		& CO(2-1) emission & -5$_{-8}^{+12}$ & 330$_{-30}^{+40}$ & 0.89$_{-0.05}^{+0.05}$ & - & - \\\rule{0pt}{3.0ex} 
		& CO(1-2) absorption `A' &237$_{-1}^{+1}$ & 17$_{-8}^{+12}$ & -2.4$_{-0.2}^{+0.2}$ & 0.29$_{-0.03}^{+0.03}$ & 4.9$_{-0.5}^{+0.6}$ \\\rule{0pt}{3.0ex} 
		& CO(1-2) absorption `B' & 269$_{-1}^{+1}$ & 21$_{-10}^{+15}$ & -1.9$_{-0.2}^{+0.2}$ & 0.23$_{-0.02}^{+0.03}$ & 4.8$_{-0.6}^{+0.7}$ \\\rule{0pt}{3.0ex} 
		& CO(1-2) absorption `C' & 336$_{-1}^{+1}$ & 8$_{-3}^{+7}$ & -2.1$_{-0.3}^{+0.4}$ & 0.24$_{-0.04}^{+0.04}$ & 2.2$_{-0.3}^{+0.4}$ \\\rule{0pt}{3.0ex} 
		& CN-A absorption & 279$_{-1}^{+1}$ & 156$_{-3}^{+3}$ & -2.40$_{-0.04}^{+0.04}$ & 0.36$_{-0.01}^{+0.01}$ & 57$_{-1}^{+1}$ \\\rule{0pt}{3.0ex} 
		& CN-B absorption & 273$_{-4}^{+4}$ & 234$_{-8}^{+8}$ & -1.03$_{-0.03}^{+0.03}$ & 0.141$_{-0.005}^{+0.005}$ & 34$_{-1}^{+1}$ \\
		\hline \hline
		RXCJ1350.3+0940 & CO(1-0) emission & -50$^{+6}_{-6}$ & 318$^{+14}_{-14}$ & 0.77$^{+0.03}_{-0.03}$ & - & - \\\rule{0pt}{3.0ex} 
		& CN-A emission & -14$^{+26}_{-26}$ & 310$^{+40}_{-50}$ & 0.19$^{+0.03}_{-0.02}$ & - & - \\\rule{0pt}{3.0ex} 
		& CN-B emission & -30$^{+30}_{-30}$ & 160$^{+50}_{-50}$ & 0.14$^{+0.04}_{-0.04}$ & - & - \\
		\hline  
		MACSJ1931.8-2634 & CO(1-0) emission & 24$^{+5}_{-6}$ & 176$^{+20}_{-15}$ & 0.66$^{+0.04}_{-0.05}$ &- & -\\ \hline
        RXCJ1603.6+1553 & CO(1-0) emission & -50$^{+7}_{-7}$ & 318$^{+18}_{-17}$ & 1.49$^{+0.07}_{-0.06}$ &- &- \\ \hline
	\end{tabular}
    \caption{The central velocity, FWHM (equivalent to 2.355 $\sigma$), amplitude, peak optical depth and velocity integrated optical depth for the absorption and emission regions shown in Fig. \ref{fig:CO_10_Absorbers_first_plot}, \ref{fig:CO_10_Absorbers_second_plot}, \ref{fig:CO_10_Absorbers_third_plot}. The velocity zero-point used for each source is given in Table \ref{tab:recession_velocities_table}. All velocities are barycentric and use the optical convention. The values and errors are calculated by performing Monte Carlo simulations which re-simulate the noise seen in each spectrum, along the same lines as described in \citet{Rose2019}. The residuals of these best fits are shown in Fig. \ref{fig:residuals_one}, \ref{fig:residuals_two} and \ref{fig:residuals_three} in Appendix \ref{sec:residuals}.}
    \label{tab:results_table}
\end{table*}

\section{Column Density Estimates}
\label{sec:ColumnDensityDerivations}

\begin{figure}
	\includegraphics[width=\columnwidth]{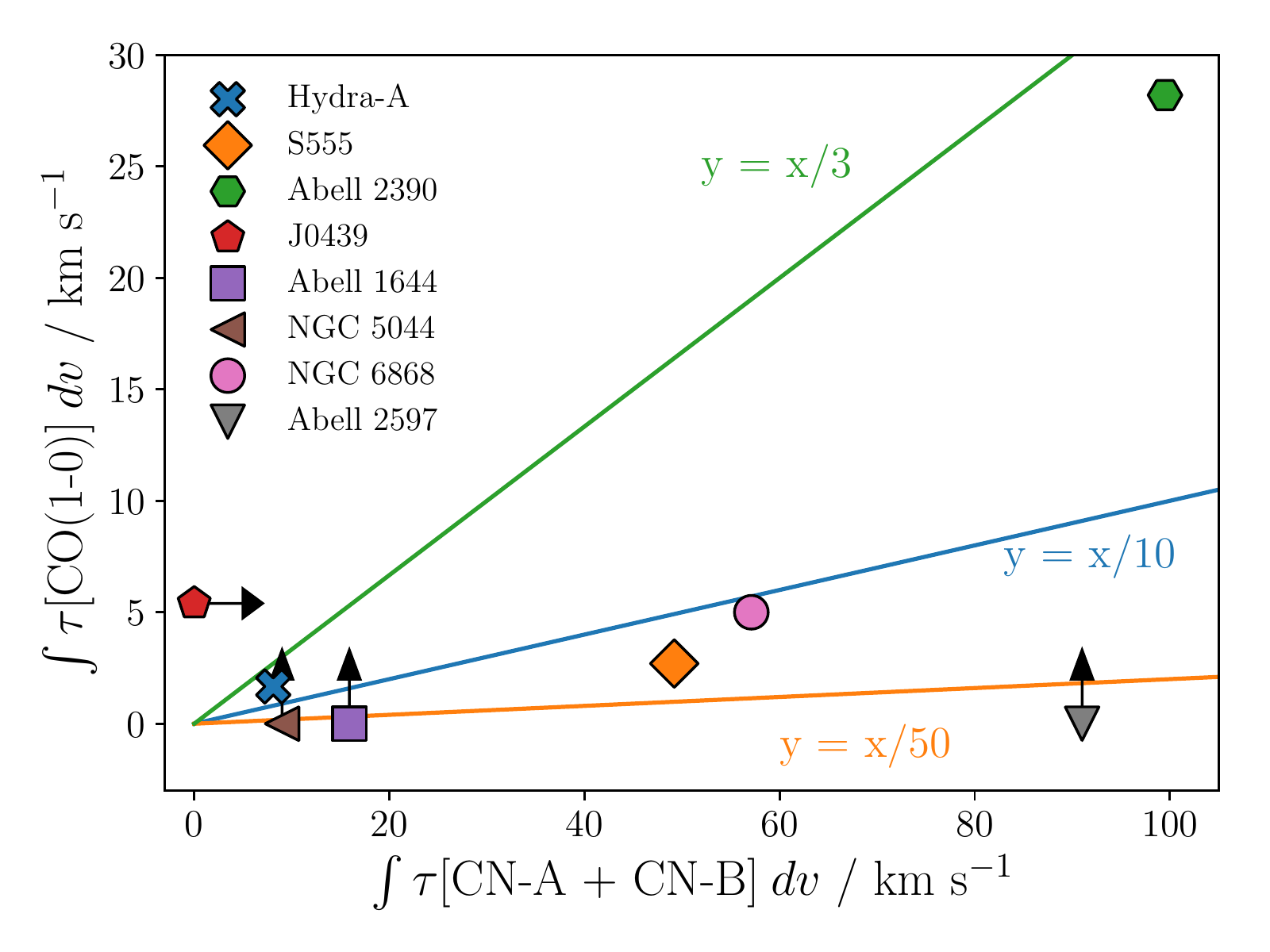}
    \caption{The velocity integrated optical depths of the CN-A + CN-B and CO(0-1) absorption lines. For most sources, the CN-A + CN-B absorption (i.e. the sum of the absorption from the various hyperfine structure lines of the $N=0-1$ transition), is typically around $\sim$10 times stronger than that of CO(0-1), indicating a molecular number ratio of CO/CN~$\sim$~10. The CN line appears stronger in absorption than that of CO despite its lower abundance because of its higher electric dipole moment.}
    \label{fig:Tau_ratio_plot}
\end{figure}

Fig. \ref{fig:Tau_ratio_plot} shows the relationship between the velocity integrated optical depths of the CO(0-1) and CN-A/CN-B lines for the eight sources in which they are detected. In the majority of cases the sum of the CN-A and CN-B absorption, i.e. the combination of all CN N = 0 - 1 hyperfine structure lines, is \mbox{$\sim$ 10} times as strong as that of CO(0-1). Using an estimated excitation temperature and treating the absorption as optically thin, it is possible to calculate the total column density, $N_{\textnormal{tot}}$, of the absorption regions, and therefore estimate the CO/CN ratio of the absorbing gas. In general,
\begin{equation}
\label{}
N_{\textnormal{tot}} = Q(T_{\textnormal{ex}}) \frac{8 \pi \nu_{ul}^{3}}{c^{3}}\frac{g_{l}}{g_{u}}\frac{1}{A_{ul}} \frac{1}{ 1 - e^{-h\nu_{ul}/k T_{\textnormal{ex}}}}\int \tau_{ul}~ dv ~,
\label{eq:colum_density}
\end{equation}
where $Q$($T_{\textnormal{ex}}$) is the partition function, $c$ is the speed of light, $A_{ul}$ is the Einstein coefficient of the observed transition and $g$ the level degeneracy, with the subscripts $u$ and $l$ representing the upper and lower levels \citep{Godard2010,Magnum2015}.

\begin{figure}
	\includegraphics[width=\columnwidth]{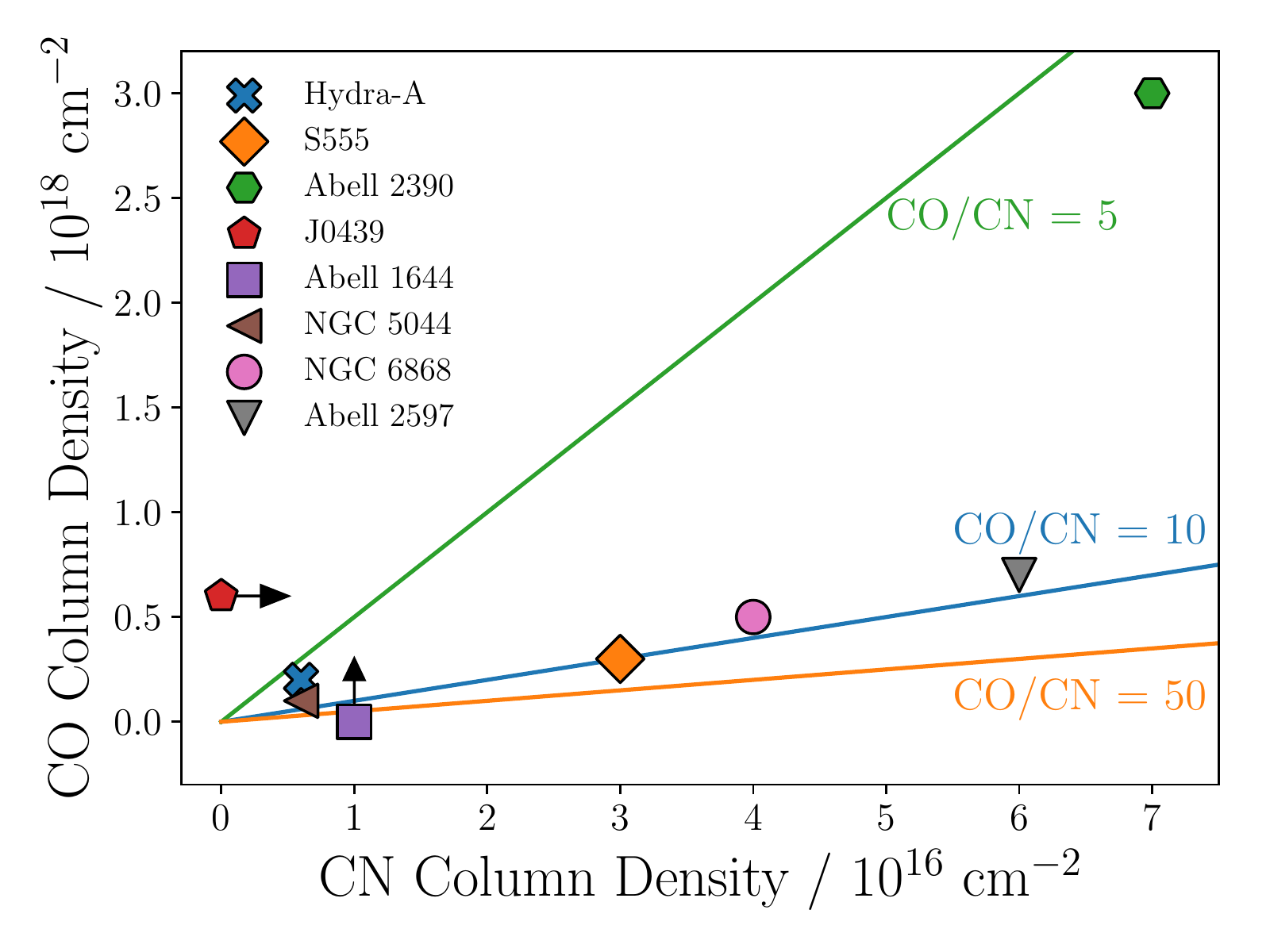}
    \caption{The total line of sight CO and CN column densities of the absorption regions shown in Fig. \ref{fig:CO_10_Absorbers_first_plot} and \ref{fig:CO_10_Absorbers_second_plot}, values of which are given in Table \ref{tab:column_densities}. These are mostly derived from the integrated optical depths shown in Table \ref{tab:results_table} and Fig. \ref{fig:Tau_ratio_plot}. However, for NGC 5044 and Abell 2597, we use archival CO(2-1) observations which show the absorption more clearly. The column densities are calculated using Eq. \ref{eq:colum_density} and assuming a gas temperature of 40 K. For most sources, the CO/CN ratio is $\sim 10$.}
    \label{fig:CN_CO_ratio_plot}
\end{figure}

The values from this calculation are given in Table \ref{tab:column_densities} and shown in Fig. \ref{fig:CN_CO_ratio_plot}. The CO/CN ratio we find for sources with both CO and CN absorption ranges from $\sim$~9 to $\sim$~44. This is similar to the values found by \cite{Wilson2018} for nearby galaxies from ALMA observations of CO and CN emission, meaning that the gas we are seeing through absorption has typical ratios of CO/CN. 

\begin{table*}
    \centering
    \begin{tabular}{lcccr}
    \hline
    Source & Temperature (K) & CO column density (cm$^{-2}$) & CN column density (cm$^{-2}$) & CO/CN ratio\\
    \hline
    Hydra-A & 20 & 5\six & 2\fif & 32$^{+4}_{-2}$ \\\rule{0pt}{3.0ex}
    & 40 & 2\sev & 6\fif & 32$^{+4}_{-2}$ \\\rule{0pt}{3.0ex}
    & 80 & 7\sev & 2\six & 32$^{+4}_{-2}$ \\
    \hline
    S555 & 20 & 8\six & 9\fif & 9$^{+2}_{-2}$ \\\rule{0pt}{3.0ex} 
    & 40 & 3\sev & 3\six & 9$^{+2}_{-2}$ \\\rule{0pt}{3.0ex}
    & 80 & 1\eig & 1\sev & 9$^{+2}_{-2}$ \\
    \hline
    Abell 2390 & 20 & 8\sev & 2\six & 44$^{+2}_{-2}$ \\\rule{0pt}{3.0ex} 
    & 40 & 3\eig & 7\six & 44$^{+2}_{-2}$ \\\rule{0pt}{3.0ex}
    & 80 & 1\nin & 3\sev & 44$^{+2}_{-2}$ \\
    \hline
    J0439+05 & 20 & 2\sev & - & - \\\rule{0pt}{3.0ex}
    & 40 & 6\sev & - & - \\\rule{0pt}{3.0ex}
    & 80 & 2\eig & - & - \\
    \hline
    Abell 1644 & 20 & - & 3\fif & - \\\rule{0pt}{3.0ex} 
    & 40 & - & 1\six & - \\\rule{0pt}{3.0ex}
    & 80 & - & 4\six & - \\
    \hline
    NGC 5044* & 20 & 4\six & 2\fif & 22$^{+2}_{-1}$ \\\rule{0pt}{3.0ex} 
    & 40 & 1\sev & 6\fif & 20$^{+2}_{-1}$ \\\rule{0pt}{3.0ex}
    & 80 & 5\sev & 2\six & 20$^{+2}_{-2}$ \\
    \hline
    NGC 6868 & 20 & 1\sev & 1\six & 14$^{+1}_{-1}$ \\\rule{0pt}{3.0ex} 
    & 40 & 5\sev & 4\six & 14$^{+1}_{-1}$ \\\rule{0pt}{3.0ex}
    & 80 & 2\eig & 1\sev & 14$^{+1}_{-1}$ \\
    \hline
    Abell 2597* & 20 & 2\sev & 2\six & 12$^{+1}_{-1}$ \\
    & 40 & 7\sev & 6\six & 11$^{+1}_{-1}$ \\\rule{0pt}{3.0ex}
    & 80 & 2\eig & 2\sev & 10$^{+1}_{-1}$ \\
    \hline
    \end{tabular}
    \caption{The CO column densities, CN column densities and molecular number ratio of CO/CN for the eight sources from Fig. \ref{fig:CO_10_Absorbers_first_plot} and \ref{fig:CO_10_Absorbers_second_plot} which have absorption regions detected. Due to its higher electric dipole moment, CN typically produces lines with a larger velocity integrated optical depth than CO despite its lower abundance. \newline *For NGC 5044 and Abell 2597, where the are no detections of CO(0-1) absorption, we use the archival CO(1-2) absorption to estimate the CO column density.}
    \label{tab:column_densities}
\end{table*}

Repeat observations of CN at high spectral resolution would be required to fully understand the relationship between CN and CO. Additionally, in the three cases where there are both \mbox{CO(1-0)} and CO(2-1) observations, the latter show stronger and clearer absorption lines. A survey of CO(2-1) is therefore vital in order to show the CO gas in more detail.

\section{Discussion}
\label{sec:Discussion}

Following the works of \citet{David2014, Tremblay2016, Ruffa2019, Rose2019}, the eight detections of molecular absorption we present significantly increases the number of brightest cluster galaxies in which cold, molecular gas has been observed in absorption against the host galaxy's bright radio core. These detections are made through CO absorption and emission lines, as well as previously undetected CN lines. In seven out of eight cases where there is a CO(0-1) detection there is also CN-A/CN-B, with the exception being RXCJ0439.0+0520. Conversely, one source, NGC 5044 shows clear CN-A/CN-B absorption, but no CO(0-1) absorption despite having been previously detected in CO(1-2) by \citet{David2014}. A weak CN line has previously been observed in the intervening absorber G0248+430 \citep{Combes2019} and was one of many lines detected in the nearby galaxy Centarus-A \citep{Ekhart1990, McCoy17}. However, these detections are notable due to their rarity, with CN absorption lines being much less commonly observed than those of CO. Further, the line has never previously been detected in absorption against a brightest cluster galaxy's bright continuum source, making our seven detections especially noteworthy.

\begin{figure}
	\includegraphics[width=0.95\columnwidth]{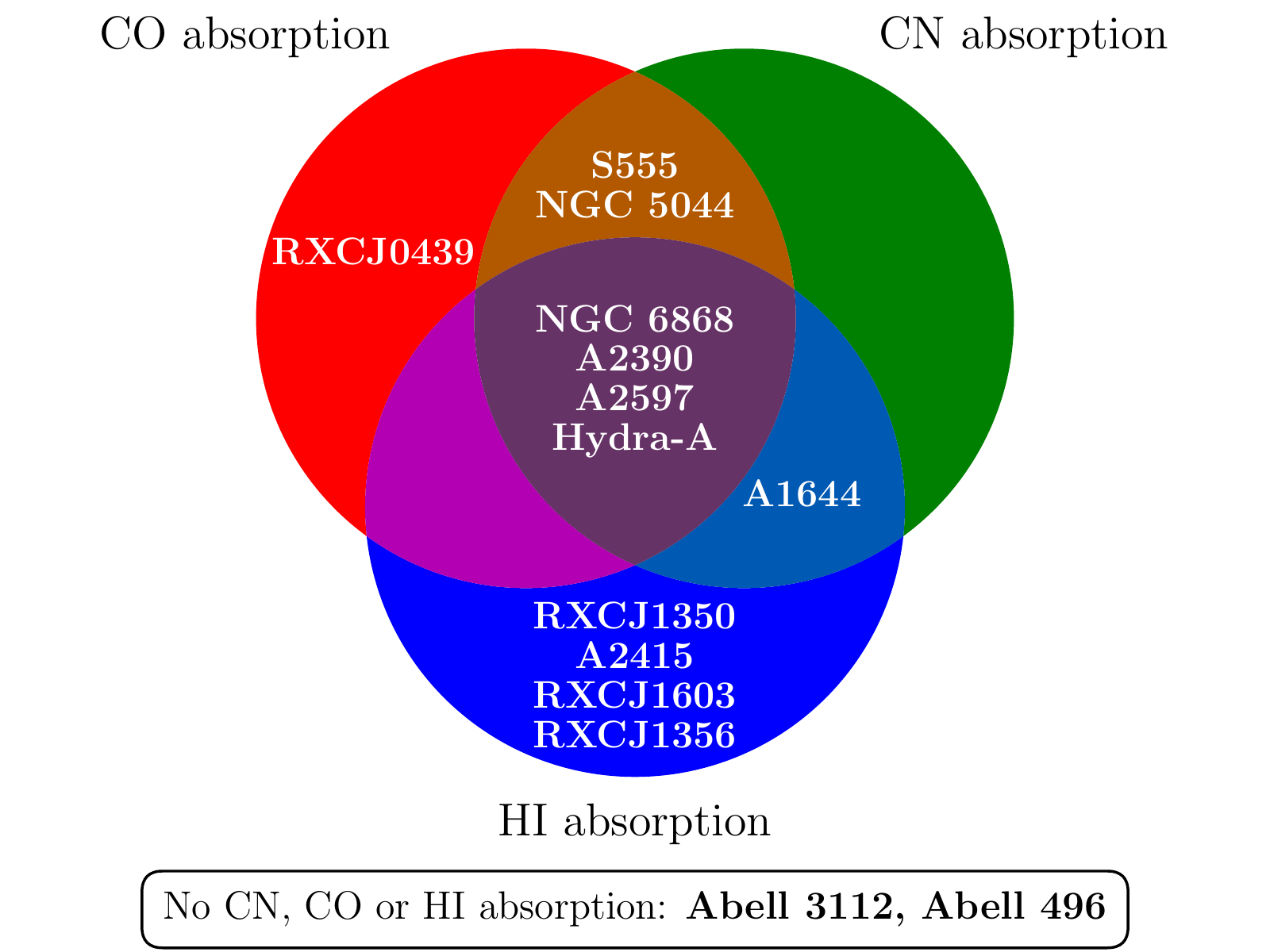}
    \caption{Venn diagram showing the combination of absorption lines which have been detected for sources which have a complete set of CO, CN and H\thinspace\small I\normalsize\space observations. Note that the CO detection of NGC 5044 has been made with the (1-2) line and there is no detection with the (0-1) line.}
    \label{fig:VennDiagram}
\end{figure}

Fig. \ref{fig:VennDiagram} shows a Venn diagram highlighting the detections of CO, CN and H\thinspace\small I\normalsize\space which have been made for sources with a complete set of observations for these lines. This emphasises the wide range in the absorption properties of these systems and implies that surveys searching for many different molecular absorption lines are justifiable, even in cases which have previous non-detections of H\thinspace\small I\normalsize\space and CO absorption.

\subsection{Potential fuelling of supermassive black holes}
\begin{figure}
	\includegraphics[width=\columnwidth]{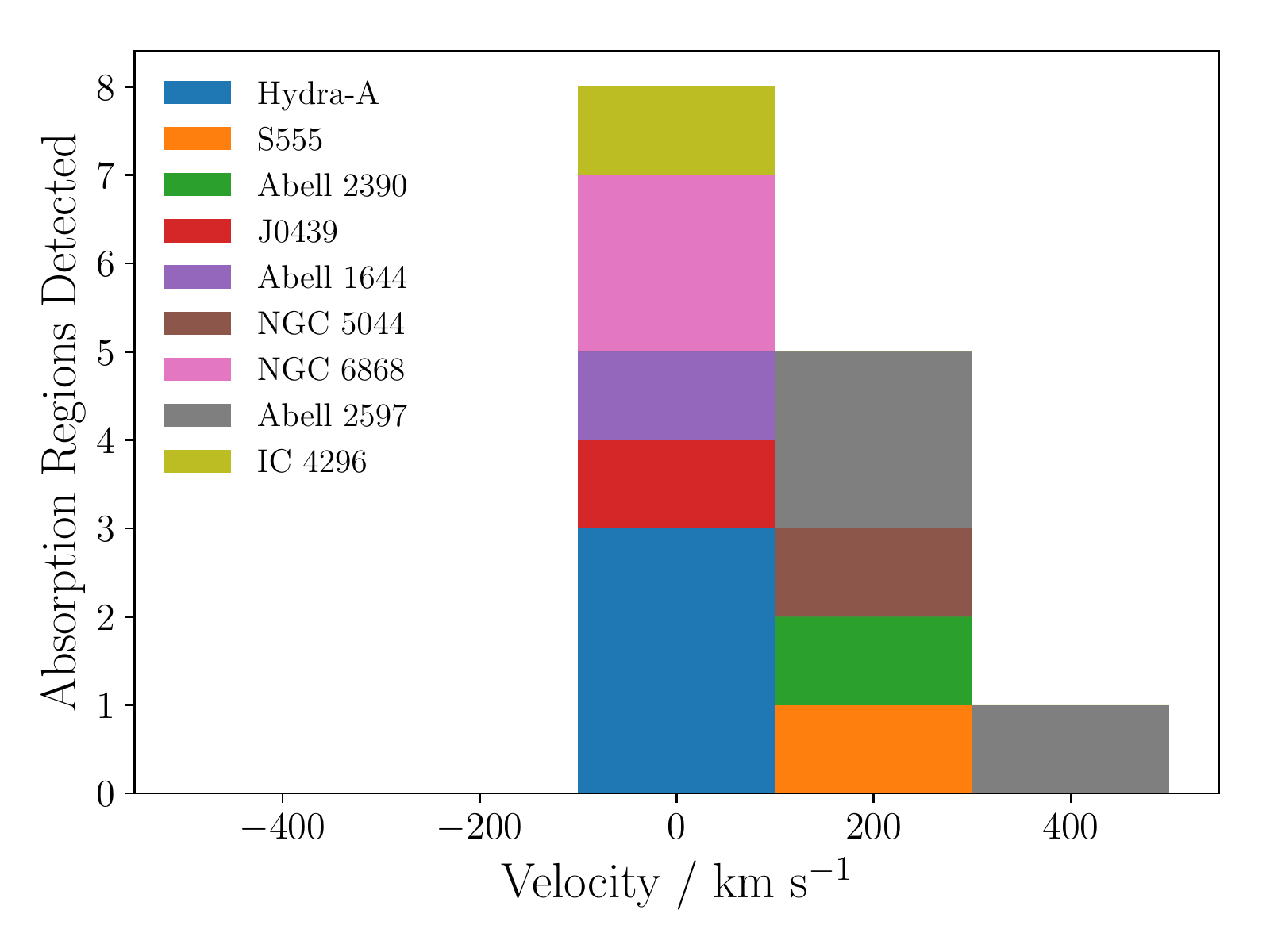}
    \caption{Histogram showing the velocities of absorbing regions detected for the nine such brightest cluster galaxy systems known to date, which has a bias for redshifted absorption. Note that from a combination of CO(0-1), CO(1-2) and CN-A/CN-B detections, we represent each galaxy's absorption feature(s) with the line which best resolves the absorption. In most cases, this is CO(0-1). However, for Abell 1644 we use CN-A and for Hydra-A, NGC 5044 and Abell 2597 we use CO(1-2). Also note that some sources have multiple absorption regions. The histogram is also unweighted by the velocity integrated optical depth of each absorption region due to the uncertainties associated with doing this for the multiple lines from different molecular species.}
    \label{fig:histogram_plot}
\end{figure}

In many cases, there is clear evidence of cold molecular gas moving towards its host galaxy's mm-continuum source at significant velocities. S555, NGC 5044 and Abell 2597 all have absorption regions with velocities towards the core of $\gtrsim 250$ \kms. Abell 2390 also has redshifted absorbing gas, albeit moving lower velocities. However, in this case the large width of the absorption implies that there is likely to be a systemic motion towards the core. Hydra-A, J0439+05, Abell 1644 and NGC 6868 all have molecular gas moving at lower blue and redshifted velocities, implying that the clouds are drifting in non-circular orbits and not experiencing any significant inflow or outflow. Overall our eight detections, combined with those of NGC 5044 \citep{David2014}, Abell 2597 \citep{Tremblay2016}, Hydra-A \citep{Rose2019} and IC 4296 \citep{Ruffa2019} do not present any evidence of significantly blueshifted absorption. Though there are some moderately blueshifted regions of molecular gas, overall there is a bias for motion towards the galaxies' supermassive black holes, as shown by Fig. \ref{fig:histogram_plot}. In the chaotic cold accretion scenario, most clouds are expected to drift in the large-scale turbulent field (with low $v_{\rm cen}$), while only a few outliers are found to reach velocities of several 100 km s$^{-1}$ \citep[see ][]{Gaspari2018}, which is consistent with our findings here. Nevertheless, the number of detections these conclusions are based upon remains small.

\subsection{Constraining the location of the absorbing clouds}

It is physically plausible that the absorbing clouds detected in these systems lie anywhere from a few tens of parsecs from the central supermassive black hole, to several kilo parsecs away. Indeed, many of the galaxies in our sample have molecular gas seen in emission out to several kpc, most notably Hydra-A where we see an edge-on gas disc with a diameter of 5 kpc. However, the covering fraction of molecular gas as a function of radius significantly constrains the distance at which the gas is likely to be, as predicted by simulations such as chaotic cold accretion \citep{Gaspari2017}. These simulations of clumpy molecular gas condensation show that the volume filling factor and internal density of molecular clouds are both inversely proportional to radius. This means that the vast majority of dense clouds which contribute to the line of sight absorption are expected to reside in the inner region, within radii of up to $\sim200$ pc (for a broader comparison of our results to chaotic cold accretion simulations, see \S\ref{sec:cca}). Conversely, two properties of the absorbing clouds imply that they lie outside the approximate Bondi capture radius in each system of a few tens of parsecs. First, the fact that the clouds are detectable by CO(0-1) absorption implies that they are all relatively cool and not being significantly heated by the high radiative power of the central AGN. Given that dust grains are found with ubiquity in interstellar gas, the approximate level of heating a molecular gas cloud will experience can be demonstrated by providing an estimate for the equilibrium dust temperature. For a dust grain radiating with a black-body spectrum, the balance between radiation and emission can be written as
\begin{equation}
    F = Q ~\sigma T^{4}~,
\end{equation}
where $F$ is the flux of the radiation field, $Q$ is the Planck average emissivity, $\sigma$ is the Stefan-Boltzmann constant and $T$ is the equilibrium temperature. Alternatively,
\begin{equation}
    \frac{L}{4\pi R^{2}} = Q ~\sigma T^{4}~,
    \label{dust_equation}
\end{equation}
where the radiation field is assumed to be from a point source of luminosity $L$ at a distance $R$. The AGN of brightest cluster galaxies such as those in our survey have typical luminosities of $10^{39} - 10^{44} \textnormal{ erg s}^{-1}$, though at the higher end these are dominated by radiatively powerful AGN \citep[see][]{Russell2013}. For a dust grain at a distance of 10 pc from a $10^{42} \textnormal{ erg s}^{-1}$ point source, the equilibrium temperature is therefore \mbox{$\sim$ 100 K}, assuming a Q value of 0.1 \citep[an approximate value from][]{Draine1984}. The existence of cold molecular gas clouds inside these distances, such as those detected in our survey, is therefore unlikely.

A second property of the absorbing clouds which implies that they lie outside the Bondi capture radius of \mbox{$\sim 10$ pc} is their velocities. Within these distances they would be expected to obtain highly redshifted velocities, perhaps of thousands of \kms, due to the gravitational influence of the central supermassive black hole. For example, in Abell 2390 which has a $3\times 10^{8}~\textnormal{M}_{\odot}$ supermassive black hole \citep{Tremblay2012}, a circular orbit at 10 pc requires a velocity of $\sim$ 400 \kms, something difficult to maintain in such a turbulent environment.

\subsection{Comparison with Chaotic Cold Accretion Simulations}
\label{sec:cca}
\begin{figure}
    \includegraphics[width=\columnwidth]{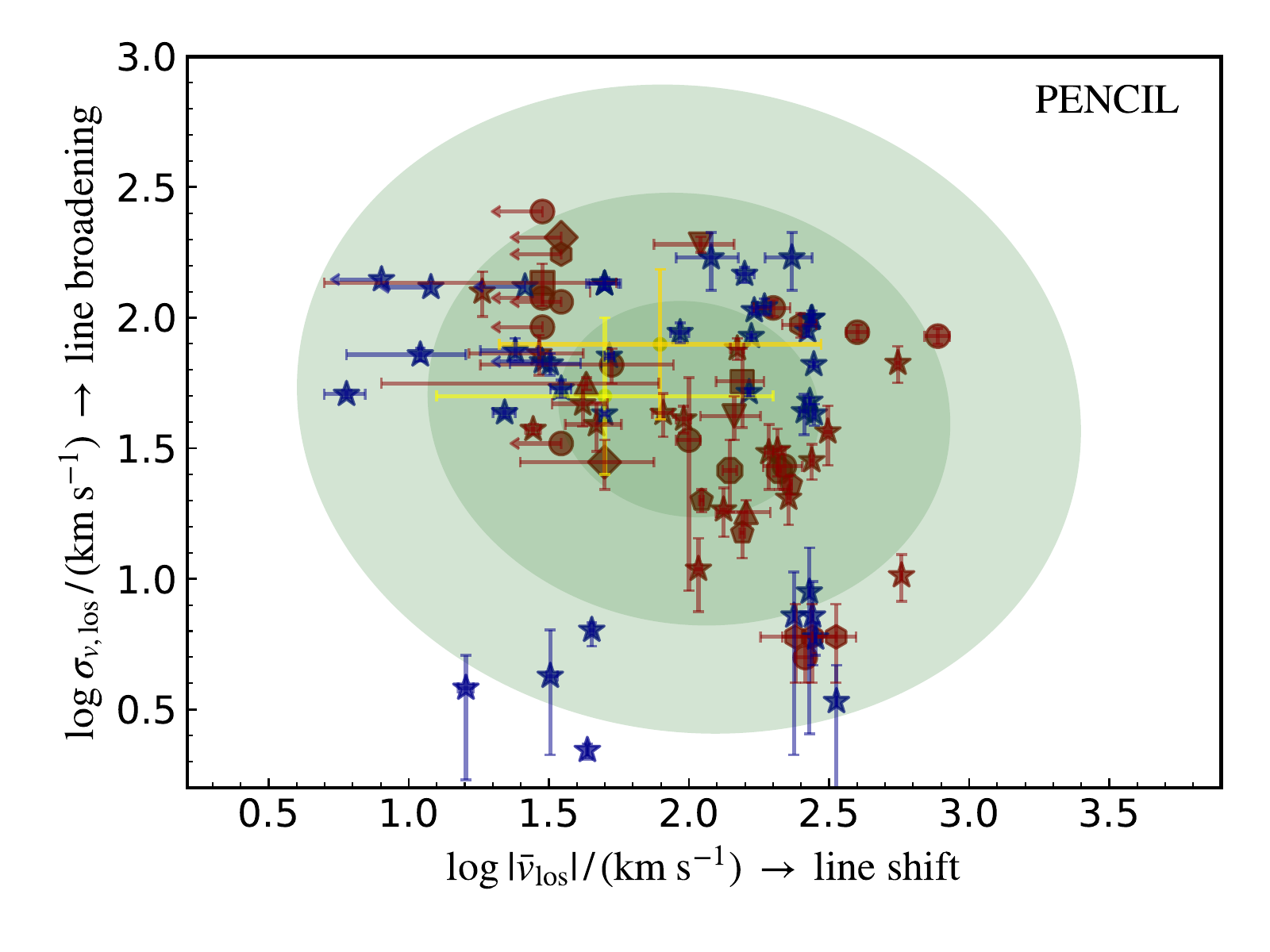}
    \caption{An analog of figure 4 from \citet{Gaspari2018}, showing the relation between the line of sight velocity dispersion (line broadening) and the magnitude of the line of sight velocity (line shift). This serves as a comparison between observational data and the predictions from chaotic cold accretion simulations ($1-3 \sigma$ green contours). The red and yellow points are observed systems with HI and CO clouds from \citet{Gaspari2018}, while the blue points show our ALMA detections from Table 5. The ALMA detections are statistically consistent with the distributions of previous data points and chaotic cold accretion predictions. Note that we include detections of the same absorption regions from different molecular tracers because they likely trace different clouds or different parts of the giant molecular associations along the line of sight. Any emission seen is highly likely to originate from large collections of clouds, though absorption may well be detected due to single clouds along the line of sight.}
    \label{fig:cca_comparison}
\end{figure}

More quantitatively, we have followed the same procedure as described in \S 4 of \citet{Gaspari2018} to compute the pencil-beam points in the main diagnostic plot of $\log \sigma_v$ versus $\log |v_{\rm shift}|$ along the line of sight to the galaxy centre. As shown in Fig. \ref{fig:cca_comparison}, the distribution of blue points (our ALMA detections in Table 5) is consistent with that of CO and HI clouds in other galaxies \citep[red and yellow; see ][]{Gaspari2018}, as well as with the simulated 1-3$\sigma$ contours predicted by chaotic cold accretion simulations. 
Regarding bulk motions, the log mean and dispersion for our points is $\log v_{\rm shift} \simeq 1.9 \pm 0.5$, which is comparable to that of the points observed in \citet{Gaspari2018} simulations with $\log v_{\rm shift} \simeq 2.0 \pm 0.5$.
In terms of the turbulence, the log mean and dispersion for our points is $\log \sigma_v \simeq 1.6 \pm 0.5$, which is analogous to that of the points observed in \citet{Gaspari2018} simulations.
It is important to note the two different classes of clouds the pencil-beam line of sight can intersect: the high-velocity single cloud (bottom) and the associations of multiple clouds that drift in the macro turbulent atmosphere (top). Interestingly, we are increasingly populating the bottom quadrants, owing to the high angular resolution of ALMA.
In future work we aim to enlarge the sample of detections to further constrain this key relationship between line broadening and velocity shift.

\subsection{Differences between the CO(1-0), CO(2-1) and CN-A/CN-B observations}

The significant differences which are seen between the strengths of the CO(1-2) absorption line detections and the lower energy CO(0-1) and CN-A/CN-B lines of NGC 5044 and Abell 2597 have a number of possible explanations. First, the CO(1-2) absorption is enhanced by a factor of three due to its statistical weight. Second, due to the time difference between the observations of CO(2-1) and CO(1-0)/CN-A/CN-B, the clouds may have moved across the line of sight to the bright radio core. The time difference between the observations is approximately 5 years in the case of NGC 5044 and 6 years for Abell 2597, whereas individual clouds are expected to take at least hundreds of years to cross the line of sight; a relatively small molecular cloud with a diameter of 0.1 pc and a large transverse velocity of 500 \kms\space will take $\sim$ 200 years to fully cross the line of sight, assuming a point-like continuum source. A third explanation is that due to the energy difference of the lines, molecular gas regions of different temperatures are being revealed by the different lines. The CO(1-2) absorption line will trace higher temperature gas than the CO(0-1) and CN-A/CN-B lines due to its higher excitation energy. Therefore, if there are multiple regions of molecular gas of significantly different temperatures along the line of sight, the lower and higher energy lines may reveal different absorption features.  However, in the case of NGC 5044, absorption is detected in the low energy CN-A/CN-B lines, but not the similarly low energy CO(0-1) line. The same absorption region is nevertheless detected in the higher energy CO(1-2) line, suggesting that whether or not absorption is present is dependent on more than just the gas temperature alone. A further factor which is likely to play a large role in affecting the strength of the absorbing regions across different lines is the molecular number ratio of CO/CN. In the case of NGC 5044, the absorption may be due to relatively warm gas with a low CO/CN ratio, resulting in modest CO(1-2) and CN-A/CN-B absorption, but no clear CO(0-1) line. 

\begin{figure*}
	\includegraphics[width=0.95\textwidth]{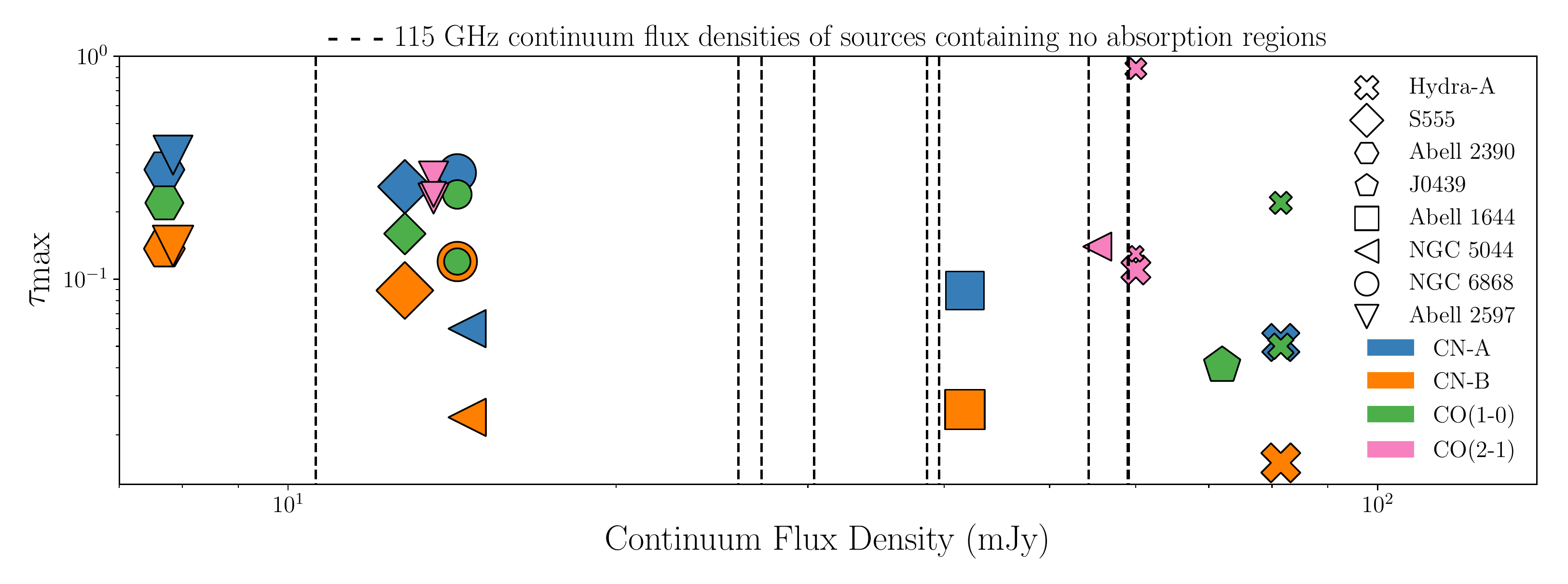}
    \caption{The relation between the continuum flux density of sources observed and the peak optical depths of their absorption regions. The approximate size of the markers is proportional to the logarithm of the absorption region's FWHM. Note that the CN-A and CN-B lines are respectively composed of five and four hyperfine structure lines and are only observed at low spectral resolution, artificially increasing their FWHM and preventing the detection of narrow lines. There is no apparent cut-off caused by the potential difficulties of detecting molecular absorption in systems with a low continuum flux density, though the narrowest absorption regions are only detected in the brightest sources. Additionally, the dashed vertical lines which mark the continuum flux densities of those sources in which no absorption lines are seen do not cluster at low flux densities, also implying that we have not reached a low brightness detection limit.}
    \label{fig:fluxdensity_v_tau_plot}
\end{figure*}

\subsection{Future observations}

In Fig. \ref{fig:fluxdensity_v_tau_plot} we show the relation between the continuum flux density of the sources observed in this survey and the peak optical depth of the absorption regions detected. This shows no obvious correlation between the continuum flux density of the sources and the number or strength of the absorption regions detected. There is also no clear cut-off as a result of potential detection limits for absorption regions in sources with low continuum flux densities. One possible exception to this is that the narrowest lines (which have the smallest markers in Fig. \ref{fig:fluxdensity_v_tau_plot}) are only seen in higher flux density systems, though only a small number of these are detected. Additionally, the systems in which we find no absorption regions, as indicated by the dashed vertical lines) have no tendency for having low flux density continuum sources. We are therefore unlikely to have met a low brightness detection limit, implying that searches for molecular absorption in lower flux density sources are justified.

ALMA Cycle 6 observations of Hydra-A, which include high spectral resolution CN N = 2 - 1 observations, show that the molecular ratio of CO/CN is a factor which can vary significantly between different absorption regions of the same system. Absorption regions `A' and `B' (see Fig. \ref{fig:CO_10_Absorbers_second_plot}) are of a similar strength in CN N = 2 - 1, despite the large difference seen in both the CO(0-1) observations and previous CO(1-2) observations \citep{Rose2019}. This implies that there is a large disparity between the CO/CN ratio across these two absorption regions and that the composite gas clouds have very different histories e.g. the CO/CN molecular number ratio can be changed if the molecular gas is present in starburst regions. These observations of Hydra-A will be presented in Rose et al. (in preparation). Detections of several other molecular species in absorption which will also be shown in Rose et al. (in preparation) also indicate that molecular line survey strategies such as those used to observe Arp 220 by \citet{Martin2011} may reap significant rewards. This includes a significant increase in our understanding of the chemical and physical properties of the molecular gas in the cores of brightest cluster galaxies as well as its origins.


\section{Conclusions}
\label{sec:conclusions}

We have presented an ALMA survey of 18 brightest cluster galaxies which lie in cool cores and have extremely bright mm-continuum sources at their centres. We find molecular absorption in eight of this sample via the detections of seven CO(0-1) absorption lines, seven CN N = 1 - 0 lines and one SiO(2-3) line, shown in Fig. \ref{fig:CO_10_Absorbers_first_plot} and \ref{fig:CO_10_Absorbers_second_plot}.

Our survey doubles the number of systems in which molecular absorption has been observed against a brightest cluster galaxy's bright continuum source from five to ten and provides new molecular absorption lines for two of those systems previously discovered.

The absorption regions we detect have velocities of between -45 to 283 \kms\space relative to the systemic recession velocity of the galaxies and overall there is a bias for motion towards the supermassive black holes, though this is found from what is still a relatively small number of sources. Our results appear to be consistent with the chaotic cold accretion scenario of \citet{Gaspari2018}. This includes the detection of drifting and infalling clouds with a covering fraction of < 1 and the statistical line broadening/shift properties of the pencil-beam diagram (Fig. \ref{fig:cca_comparison}).

Given that we find eight absorbing systems from the observed sample of 18, it is highly unlikely that a detection rate this high could be produced by absorption at large distances. Instead, we have most likely found cases of absorption due to molecular gas at distances within which they could feasibly be accreted onto the supermassive black hole under the right conditions. At these distances of up to a few hundred parsecs, slightly elliptical orbits would be expected to produce offsets of just a few tens of \kms, rather than the hundreds of \kms\space we see in some of our observations i.e. these large velocities relative to the galaxies' recession velocities are not due to orientation effects.

We find that CN is a significantly stronger tracer of molecular absorption than CO due to the molecule's higher electric dipole moment. From the eight sources which have detections of both lines, the velocity integrated optical depths are $\sim$ 10 times higher for CN. This implies a typical molecular number ratio of CO/CN~$\sim$~10.

The CO(1-2) line also appears to be a more efficient tracer of molecular absorption than the lower energy CO(0-1) line. Observations of both lines now exist for three sources: Hydra-A, Abell 2597 and NGC 5044. In all cases, the absorption features appear significantly deeper and clearer in the higher energy line.

With the additions of our survey, a complete set of CO, CN and H\thinspace\small I\normalsize\space observations now exists for 14 sources (Fig. \ref{fig:VennDiagram}). From these, many different combinations of absorption lines are detected. For four sources, all three lines are detected while for a further four only H\thinspace\small I\normalsize\space absorption is seen. Two show both CO and CN absorption but not that of H\thinspace\small I\normalsize\space. One source shows only CO absorption while another shows both CN and H\thinspace\small I\normalsize\space absorption but not that of CO. For two sources, none of the three absorption lines are seen. In relation to future surveys, these results imply that non-detections of a particular absorption line do not rule out subsequent detections of other lines.

\section*{Acknowledgements}

The authors gratefully acknowledge the anonymous referee for their comments, which helped us to improve the paper.

We thank Tom Oosterloo for generously providing the H\thinspace\small I\normalsize\space detection of NGC 6868. 

T.R. is supported by the Science and Technology Facilities Council (STFC) through grant ST/R504725/1.

A.C.E. acknowledges support from STFC grant ST/P00541/1.

M.G. is supported by the Lyman Spitzer Jr. Fellowship
(Princeton University) and by NASA Chandra grants GO7-18121X and GO8-19104X.

S.B. and C.O. are grateful for support from the Natural Sciences and Engineering Research Council of Canada.

G.R.T. acknowledges support from the National Aeronautics and Space Administration (NASA) through Chandra Award Number GO7-8128X8, issued by the Chandra X-ray Center, which is operated by the Smithsonian Astrophysical Observatory for and on behalf of NASA under contract NAS8-03060.

This paper makes use of the following ALMA data: ADS/JAO.ALMA\#2017.1.00629.S. ALMA is a partnership of ESO (representing its member states), NSF (USA) and NINS (Japan), together with NRC (Canada), NSC and ASIAA (Taiwan), and KASI (Republic of Korea), in cooperation with the Republic of Chile. The Joint ALMA Observatory is operated by ESO, AUI/NRAO and NAOJ. We aslo use archival data: ADS/JAO.ALMA\#2016.1.00533.S of NGC 5044 and of Abell 2597.






\section*{List of institutions}
\textit{
\noindent $^{1}$Centre for Extragalactic Astronomy, Durham University, DH1 3LE, UK\\
$^{2}$LERMA, Observatoire de Paris, PSL Research Univ., College de France, CNRS, Sorbonne Univ., Paris, France\\
$^{3}$Department of Astrophysical Sciences, 4 Ivy Lane, Princeton University, Princeton, NJ 08544-1001, USA\\
$^{4}$ Department of Physics, University of Bath, North Rd, Bath, BA2 7AY \\
$^{5}$Institut d'Astrophysique Spatiale, Centre Universitaire d'Orsay, 91405 Orsay, France\\
$^{6}$Gemini Observatory, Northern Operation Center, 67-0 N. A'Ohoku Place, Hilo, HI, USA \\
$^{7}$Department of Astronomy, University of Virginia, 530 McCormick Road, Charlottesville, VA 22904-4325, USA\\
$^{8}$Harvard-Smithsonian Center for Astrophysics, 60 Garden St., Cambridge, MA 02138, USA\\
$^{9}$Department of Physics \& Astronomy, University of Manitoba, Winnipeg, MB R3T 2N2, Canada \\
$^{10}$Chester F. Carlson Center for Imaging Science, Rochester Institute of Technology, 84 Lomb Memorial Dr., NY 14623, USA\\
$^{11}$HH Wills Physics Laboratory, Tyndall Avenue, Bristol, BS8 1TL, UK\\
$^{12}$Department of Physics and Astronomy, University of Waterloo, Waterloo, ON N2L 3G1, Canada\\
$^{13}$School of Physics and Astronomy, Rochester Institute of Technology, 85 Lomb Memorial Drive, USA\\
$^{14}$SURFsara, P.O. Box 94613, 1090 GP Amsterdam, The Netherlands\\
$^{15}$ASTRON, Netherlands Institute for Radio Astronomy, 7990AA Dwingeloo, The Netherlands\\
$^{16}$Leiden Observatory, Leiden University, Niels Borhweg 2, NL-2333 CA Leiden, The Netherlands\\
$^{17}$Institute of Astronomy, Cambridge University, Madingly Rd., Cambridge, CB3 0HA, UK\\
$^{18}$Physics \& Astronomy Department, Michigan State University, East Lansing, MI 48824-2320, USA\\
$^{19}$Department of Physics and Astronomy, University of Kentucky, Lexington, Kentucky 40506-0055, USA\\
$^{20}$RIT College of Science, 85 Lomb Memorial Drive, Rochester, NY 14623, USA\\}


\appendix
\section{Observation details for sources lacking emission and absorption lines}
\label{sec:appendix,nondetections}

Table \ref{tab:observations_table_non_detections} shows details of the observations for which no $\geq 3\sigma$ detections of emission or absorption were made with either the CO(1-0), CN-A or CN-B lines.
\begin{table*}
	\begin{tabular}{lcccr} 

		\hline
		 & MACSJ0242.5-2132 & Abell 3112 & Abell 496 & RXCJ0132.6-0804 \\
		\hline
		Observation date & 2018 Jan 12 & 2018 Jan 11 & 2018 Jan 13 & 2018 Jan 16 \\
		Integration time (s) & 1300 & 1300 & 6800 & 2500 \\
		CO(1-0) vel. resolution (\kms) & 3.3 & 2.7 & 2.6 & 2.9 \\
		Frequency resolution (kHz) & 977 & 977 & 977 & 977 \\
		Angular resolution (arcsec) & 0.48 & 0.54 & 0.48 & 0.62 \\
		PWV (mm) & 6.5 & 6.6 & 2.2 & 4.3 \\
		FoV (arcsec) & 67.9 & 62.9 & 60.2 & 67.5 \\
		ALMA configuration & C43-5 & C43-5 & C43-5 & C43-5 \\
		Maximum spacing (m) & 1400 & 1400 & 1400 & 1400 \\
		CO(1-0) noise per channel (mJy) & 1.87 & 1.10 & 0.55 & 1.80 \\
		115 GHz cont. flux density (mJy) & 39.6 & 30.4* & 59.0 & 38.6 \\
		\hline
	\end{tabular}

	\begin{tabular}{lccr} 

		\hline
		 & Abell 2415 & Abell 3581 & RXCJ1356.0-3421 \\
		\hline
		Observation date & 2018 Jan 23 & 2018 Sep 11 & 2018 Sep 11 \\
		Integration time (s) & 5400 & 1700 & 1200 \\
		CO(1-0) vel. resolution (\kms) & 2.7 & 2.6 & 3.1 \\
		Frequency resolution (kHz) & 977 & 977 & 977 \\
		Angular resolution (arcsec) & 0.57 & 0.92 & 0.91 \\
		PWV (mm) & 1.76 & 0.72 & 0.7 \\
		FoV (arcsec) & 61.9 & 59.5 & 63.4 \\
		ALMA configuration & C43-5 & C43-4 & C43-4 \\
		Maximum spacing (m) & 1400 & 784 & 784 \\
		CO(1-0) noise per channel (mJy) & 0.35 & 0.85 & 1.32 \\
		115 GHz cont. flux density (mJy) & 27.2 & 59.1 & 25.9  \\
		\hline
	\end{tabular}

    \caption{A summary of the ALMA observations in which no absorption or emission lines were detected, all of which were taken using ALMA band 3. *The continuum source of Abell 3112 is extended to the North West, with a flux density peaking at 12.9 mJy.}
    \label{tab:observations_table_non_detections}
\end{table*}

\section{Residuals of Absorption and Emission Line Fits}
\label{sec:residuals}

\begin{figure*}
    \includegraphics[width=\textwidth]{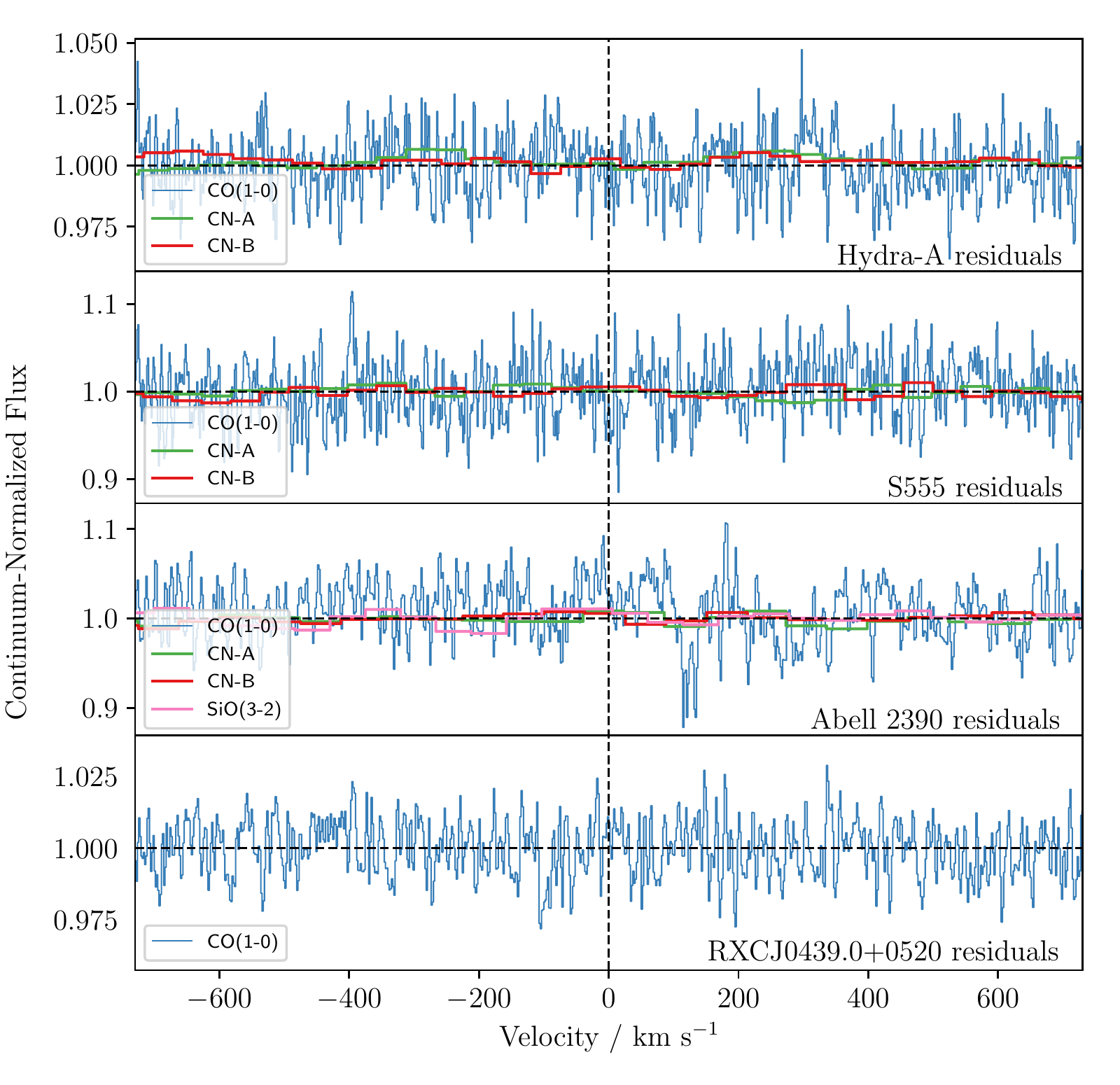}
    \caption{Fig. \ref{fig:residuals_one} and \ref{fig:residuals_two} are the residuals for the spectra shown in Fig. \ref{fig:CO_10_Absorbers_first_plot} and \ref{fig:CO_10_Absorbers_second_plot} and their Gaussian best fits as given in Table \ref{tab:results_table}. These are the sources which have CO and/or CN emission along the line-of-sight to their bright continuum sources. The only residual plot which reveals detail not encapsulated by the Gaussian best fit is the CO(1-0) spectrum of Abell 2390. This is due to the saw tooth shape of the absorption (see Fig. \ref{fig:CO_10_Absorbers_first_plot}).}
    \label{fig:residuals_one}
\end{figure*}

\begin{figure*}
    \includegraphics[width=\textwidth]{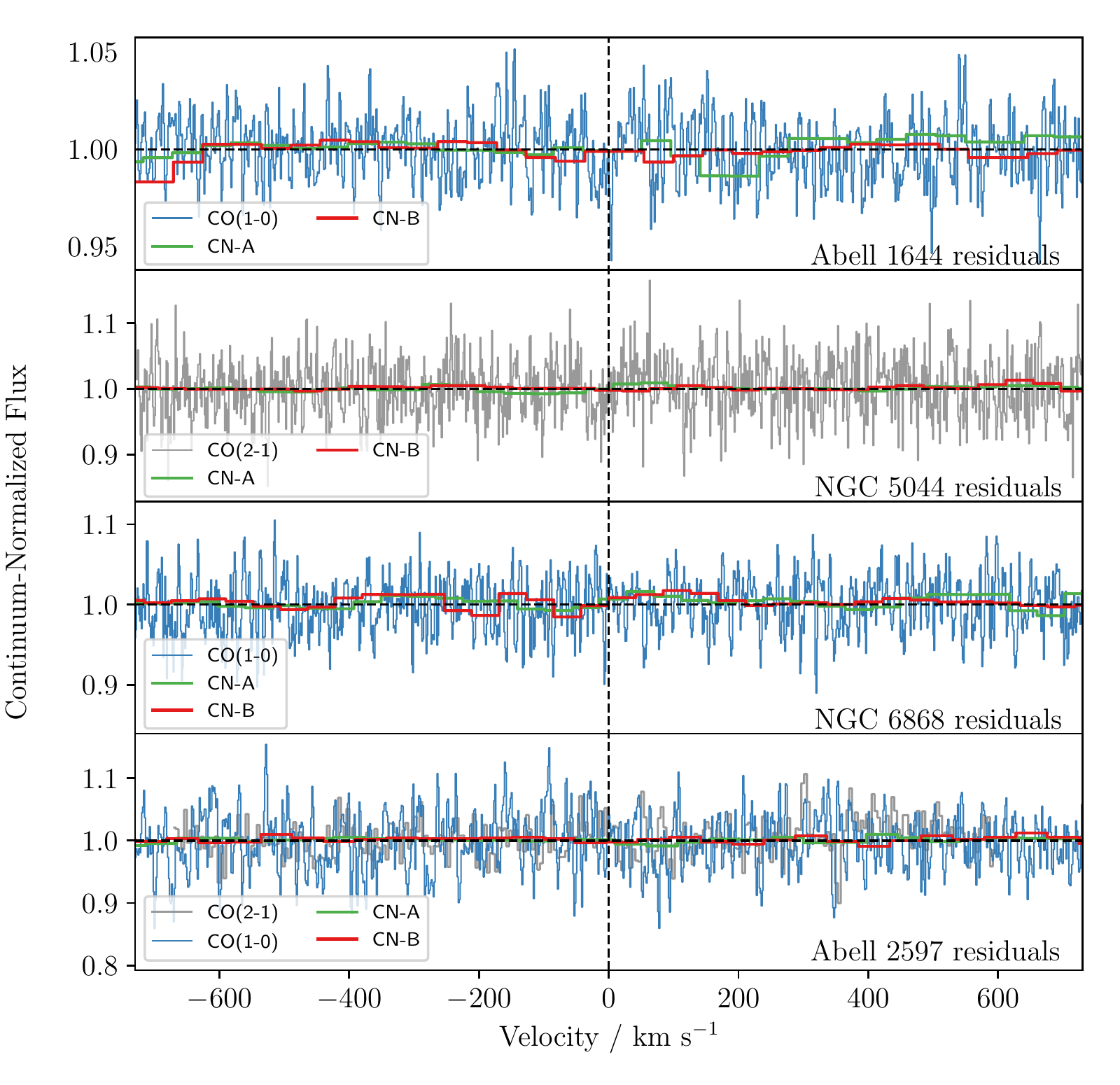}
    \caption{Residuals for the spectra shown in Fig. \ref{fig:CO_10_Absorbers_first_plot} and their Gaussian best fits as given in Table \ref{tab:results_table}. Continued from Fig. \ref{fig:residuals_one}.}
    \label{fig:residuals_two}
\end{figure*}

\begin{figure*}
    \includegraphics[width=\textwidth]{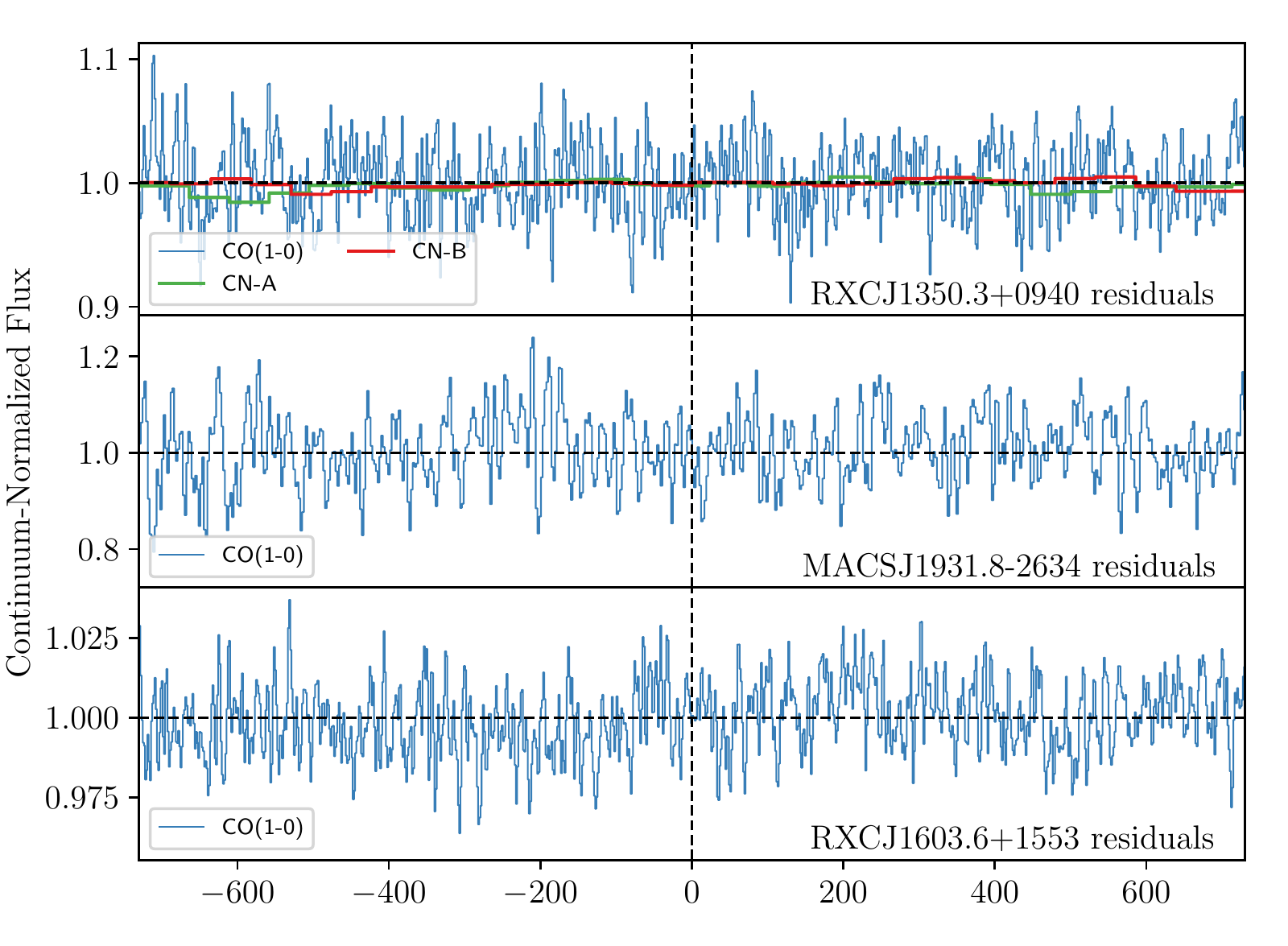}
    \caption{Residuals for spectra shown in Fig. \ref{fig:CO_10_Absorbers_third_plot}, calculated with their best fits as given in Table \ref{tab:results_table}. These are the sources which have CO and/or CN emission, but not absorption.}
    \label{fig:residuals_three}
\end{figure*}

Fig. \ref{fig:residuals_one}, \ref{fig:residuals_two} and \ref{fig:residuals_three} show the residuals for the spectra shown in Fig. \ref{fig:CO_10_Absorbers_first_plot}, \ref{fig:CO_10_Absorbers_second_plot} and \ref{fig:CO_10_Absorbers_third_plot}, calculated using the Gaussian best fits for the emission and absorption lines which are given in Tab. \ref{tab:results_table}.

\bsp	
\label{lastpage}
\end{document}